\documentclass[aps,nofootinbib,floatfix,showpacs,preprintnumbers,twocolumn]{revtex4} %eqsecnum,
\usepackage{graphicx}
\usepackage{bm}
\usepackage{mathrsfs}
\usepackage{float}
\usepackage{tabularx} 
\usepackage{amsmath}
\usepackage{epstopdf}
\usepackage{color}
\usepackage{hyperref}
\input epsf

\def\lsim{\mathrel{\raise.3ex\hbox{$<$\kern-.75em\lower1ex\hbox{$\sim$}}}}
\def\gsim{\mathrel{\raise.3ex\hbox{$>$\kern-.75em\lower1ex\hbox{$\sim$}}}}

\def\cmm2{{\,\rm cm^{-2}}}
\def\cm2{{\,{\rm cm}^2}}
\def\cmm3{{\,{\rm cm}^{-3}}}
\def\gcmm3{{\,{\rm g\,cm^{-3}}}}

\def\fun#1#2{\lower3.6pt\vbox{\baselineskip0pt\lineskip.9pt
  \ialign{$\mathsurround=0pt#1\hfil##\hfil$\crcr#2\crcr\sim\crcr}}}

\def\etal{{\it et al.}}

\def\be{\begin{equation}}
\def\ee{\end{equation}}
\def\bea{\begin{eqnarray}}
\def\eea{\end{eqnarray}}

\begin{document}
%\modulolinenumbers[5]
%\linenumbers
%\preprint{FERMILAB-PUB-09-346-A-PPD, MAN/HEP/2009/31}

%\vspace{.2in}
\title{
Readout strategies for directional dark matter detection beyond the neutrino background}

\author{Ciaran A. J. O'Hare }\email{ciaran.ohare@nottingham.ac.uk} \affiliation{School of Physics and Astronomy, University of Nottingham, University Park, Nottingham, NG7 2RD, UK}

\author{Anne M. Green}\affiliation{School of Physics and Astronomy, University of Nottingham, University Park, Nottingham, NG7 2RD, UK}

\author{Julien Billard} \affiliation{IPNL, Universit\'e de Lyon, Universit\'e Lyon 1, CNRS/IN2P3, 4 rue E. Fermi 69622 Villeurbanne cedex, France}

\author{Enectali Figueroa-Feliciano} \affiliation{Department of Physics, Massachusetts Institute of Technology, Cambridge, MA 02139, USA}

\author{Louis E. Strigari} \affiliation{Mitchell Institute for Fundamental Physics and Astronomy, 
Department of Physics and Astronomy, Texas A \& M University, College Station, TX 77843-4242, USA}

\date{\today}
\smallskip
\begin{abstract}
The search for weakly interacting massive particles (WIMPs) by direct detection faces an encroaching background due to coherent neutrino-nucleus scattering. As the sensitivity of these experiments improves, the question of how to best distinguish a dark matter signal from neutrinos will become increasingly important. A proposed method of overcoming this so-called ``neutrino floor'' is to utilize the directional signature that both neutrino and dark matter induced recoils possess. We show that directional experiments can indeed probe WIMP-nucleon cross-sections below the neutrino floor with little loss in sensitivity due to the neutrino background. In particular we find at low WIMP masses (around 6 GeV) the discovery limits for directional detectors penetrate below the non-directional limit by several orders of magnitude. For high WIMP masses (around 100 GeV), the non-directional limit is overcome by a factor of a few. Furthermore we show that even for directional detectors which can only measure 1- or 2-dimensional projections of the 3-dimensional recoil track, the discovery potential is only reduced by a factor of 3 at most. We also demonstrate that while the experimental limitations of directional detectors, such as sense recognition and finite angular resolution, have a detrimental effect on the discovery limits, it is still possible to overcome the ultimate neutrino background faced by non-directional detectors.

\end{abstract}
\maketitle

\section{Introduction}
\label{sec:intro}

Various cosmological observations indicate that $\sim 30 \%$ of the energy density of the Universe is in the form of cold, non-baryonic, dark matter (CDM)~\cite{Ade:2015xua}. Weakly Interacting Massive Particles (WIMPs) are a good CDM candidate; they arise in extensions of the Standard Model of Particle Physics, such as Supersymmetry, and are naturally produced in the early Universe with the correct abundance (for reviews see e.g., Refs.~\cite{Jungman,Bertone}). WIMPs from the dark matter halo of the Milky Way can be detected directly on Earth, via the keV-scale recoils produced when they elastically scatter off nuclei~\cite{Goodman:1984dc}.

Current direct detection experiments, including CDMS~\cite{Agnese:2014aze}, LUX~\cite{Akerib:2013tjd} and Xenon100~\cite{Aprile:2011hx}, are sensitive to spin-independent WIMP-nucleon cross-sections of order $\sigma_{\chi-n} \approx  10^{-44}-10^{-45} \, {\rm cm}^2$.
Significant increases in sensitivity are expected in the next few years as detector target masses are increased to the ton-scale and beyond~(see e.g., Ref.~\cite{baudisrev}). As anticipated in early work on direct detection~\cite{annual}, these large detectors will also be able to detect coherent neutrino-nucleus scattering of astrophysical neutrinos~\cite{Cabrera:1984rr,neutrinoJocelyn,Strigari:2009bq,Gutlein:2010tq}. Neutrinos are therefore the ultimate background for WIMP direct detection searches as they cannot be shielded against and produce recoils with similar rates and energy spectra~\cite{neutrinoJocelyn,Strigari:2009bq,Gutlein:2010tq,neutrinoBillard}.

For near-future direct detection experiments the most problematic types of neutrino are those  produced in $^8$B decay in the Sun and in cosmic ray collisions in the Earth's atmosphere. In a Xenon detector the recoil energy spectrum and rate from ${}^{8}\rm{B}$ neutrinos very closely matches that of a WIMP with mass $m_{\chi}= 6 \, {\rm GeV}$ and cross-section $\sigma_{\chi-n} \sim 5 \times 10^{-45} \, {\rm cm}^2$, while the spectrum from atmospheric neutrinos is similar to that of a WIMP with $m_{\chi} \sim 100 \, {\rm GeV}$ and  $\sigma_{\chi-n} \sim  10^{-48} \, {\rm cm}^2$~\cite{Strigari:2009bq}. Consequently the sensitivity of an experiment to WIMPs reaches a point of saturation where it becomes difficult to tell the difference between WIMP and neutrino induced recoils using their energies alone. So as the exposure or mass of an experiment increases the minimum discoverable cross-section rather than decreasing reaches a plateau. The value of cross-section at which this occurs depends on the systematic uncertainty in the neutrino flux and is commonly referred to as the ``neutrino floor''~\cite{neutrinoBillard}. It is worth noting however that for very large exposures, the neutrino floor is eventually mitigated when the small differences in the tails of the recoil energy distributions of WIMPs and neutrinos start to distinguish the two. However these required exposures are prohibitively large and well beyond the next generation of experiments so for practical purposes the neutrino background does effectively present a ``floor'' to the discovery of dark matter and we will refer to it as such.

If we wish to probe cross-sections below the neutrino floor then it is crucial to search for ways to distinguish the WIMP and neutrino signals, for instance via their different time and direction dependences. Grothaus \etal~\cite{Grothaus:2014hja} have explored the sensitivity of directional detectors capable of measuring the recoil directions in 3-dimensions using a hypothesis test which fits the WIMP and neutrino event rates as a function of energy, time, and event angle with respect to the Earth-Sun direction. Davis \cite{Davis:2014ama} found that with very large exposures adding timing information allows the neutrino floor to be evaded at low WIMP masses, due to the (small) annual modulation of both the WIMP and Solar neutrino signals. Ruppin \etal~\cite{neutrinoRuppin} examined how combining data from detectors composed of different target materials can probe cross-sections below the neutrino floor of a single experiment. They found that for spin-independent interactions the similarity in how the WIMP and neutrino signals scaled with respect to different target nuclei limited the advantage gained from multiple experiments. However for spin-dependent interactions the complementarity of multiple targets greatly improves the potential discovery limits.

Directional detection experiments aim to reconstruct the nuclear recoil tracks in 3-dimensions. However this is experimentally challenging, and greater sensitivity might be achieved with a larger detector which measures the 1- or 2-dimensional projection of the recoil tracks. In this paper we extend the work of Ref.~\cite{Grothaus:2014hja} by studying the effect of neutrino backgrounds on the sensitivity of ideal directional detectors with 1-, 2- and 3-dimensional readout. We compare the potential discovery limits of these detectors with those of non-directional experiments that only measure the energy of the recoils, or only count the number of events above some threshold energy. We also extend our study beyond the ideal detector case and consider the effects of finite angular resolution and limited sense recognition.

The paper is organized as follows. In Section~\ref{sec:NeutrinoFluxes} we discuss the direction dependence of the cosmic neutrino fluxes. We then in Section~\ref{sec:rates} use these fluxes to calculate the directional neutrino event rates, and also review the calculation of the WIMP directional event rate. In Section~\ref{sec:analysis} we outline our analysis methodology for calculating discovery limits. We present our results in Section~\ref{sec:res}. Finally we conclude in Section~\ref{sec:conc} with a summary and discussion of our results.

%------------------------------------------------------------
%                           NeutrinoFluxes
%------------------------------------------------------------

\section{Neutrino Fluxes}
\label{sec:NeutrinoFluxes}
\par In this section we review the Solar, atmospheric and diffuse supernovae background neutrino (DSNB) fluxes that dark matter detectors will be sensitive to. We expand upon previous results in the literature by highlighting the angular dependence of these fluxes. 

\subsection{Solar neutrinos}

\par
\begin{figure}[t]
\begin{center}
\includegraphics[trim = 5mm 0 0mm 3mm, clip, width=0.49\textwidth,angle=0]{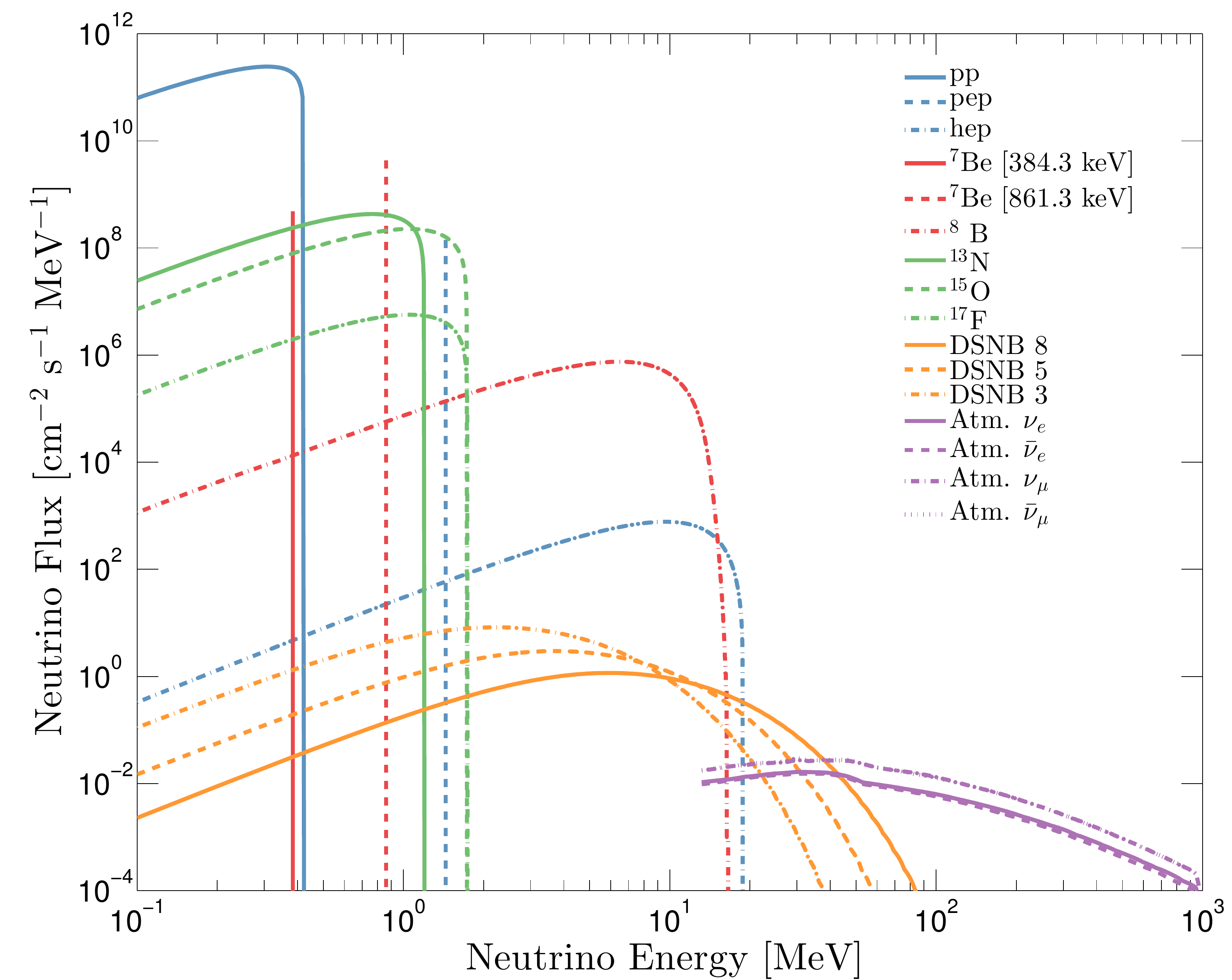}
\caption{Neutrino energy spectra which are backgrounds to direct detection experiments: Solar, atmospheric, and the diffuse supernovae background. The Solar neutrino fluxes are normalized to the high metallicity SSM. The atmospheric neutrinos are split into electron, antielectron, muon and antimuon neutrino components. The three DSNB spectra are labelled by their temperature in MeV, see Sec.\ref{sec:dsnb}.} 
\label{fig:Flux}
\end{center}
\end{figure}

 Neutrinos produced from several reactions in the Solar interior have been well-measured (for recent reviews see Refs.~\cite{Robertson:2012ib,Antonelli:2012qu}). Most recently, the Borexino experiment has made the first spectral measurement of the $pp$ component of the Solar neutrino flux~\cite{Bellini:2014uqa}. The theoretical systematic uncertainties on different components of the Solar neutrino flux range from 1\% ($pp$ flux) to 14\% ($^8$B flux). For all measurements except the $pp$ component, the theoretical uncertainties are as large as or larger than the measurement uncertainties. The theoretical uncertainty arises largely from the uncertainty in the Solar metallicity, and in order to establish a self-consistent model of Solar neutrino fluxes one must assume a metallicity model.  For the Solar neutrino flux we utilize the high metallicity Standard Solar Model (SSM) as defined in Ref.~\cite{Robertson:2012ib}; low metallicity SSMs can predict lower flux normalizations by up to $\sim 14$\% depending on the flux component. We consider here the high metallicity SSM because at present this model is more consistent with both the SNO neutral current measurement and heliosiesmology data. In fact, future dark matter detection experiments will shed further light on the Solar metallicity issue~\cite{Billard:2014yka}. 
Figure~\ref{fig:Flux} shows the fluxes and normalizations of neutrinos produced from the different reactions. Due to their rather low energy ($\mathcal{O}(1-10)$ MeV) compared to the DSNB and atmospheric neutrinos, Solar neutrinos will mostly impact the discovery potential of future direct detection experiments in the low-mass region, below 10~GeV (see Ref.~\cite{neutrinoBillard,neutrinoRuppin}).

\par In addition to the overall fluxes of Solar neutrinos, we are interested in their direction and time dependence. Due to the eccentricity of the Earth's orbit, the Earth-Sun distance has an annual variation inducing a modulation in the Solar neutrino flux as seen by an Earth-based experiment such that,
\begin{align}
  \frac{\textrm{d}^3 \Phi}{\textrm{d}E_\nu \textrm{d}\Omega_\nu \textrm{d}t}  = & \frac{\textrm{d} \Phi}{\textrm{d} E_\nu} \, \times \frac{1}{\Delta t}\left[ 1 + 2\epsilon\cos\left(\frac{2\pi(t- t_\nu)}{T_\nu}\right) \right]\nonumber \\
& \times \delta\left(\hat{{\bf q}}_\nu-\hat{{\bf q}}_\odot(t)\right) \,,
\label{eq:solarneutrinoflux}
\end{align}
where $t$ is the time from January 1st, $\epsilon = 0.016722$ is the eccentricity of the Earth's orbit, $t_{\nu} = 3$ days is the time at which the Earth-Sun distance is minimum (and hence the Solar neutrino flux is largest), $T_{\nu} = 1$ year, $\Delta t$ is the duration of the measurement, $\hat{{\bf q}}_{\nu}$ is a unit vector in the direction of interest and $\hat{{\bf q}}_\odot(t)$ is a unit vector in the inverse of the direction of the Sun\footnote{We ignore the angular size of the Sun's core on the sky which would give a tiny angular spread in the incoming neutrino directions}. As shown in Ref.~\cite{Davis:2014ama}, both the Solar neutrino and WIMP event rates have a $\sim 5$\% annual modulation but they peak at times that are separated by about 5 months, and consequently timing information could help discriminate WIMPs from neutrinos.

\subsection{Atmospheric neutrinos}
\par At higher nuclear recoil energies, greater than approximately 20 keV, the neutrino floor at high WIMP masses, i.e., above 100 GeV, will mostly be induced by low-energy atmospheric neutrinos (see~\cite{neutrinoBillard,neutrinoRuppin}). These will limit the sensitivity of dark matter detectors without directional sensitivity to spin independent cross-sections greater than approximately $10^{-48}$ cm$^2$~\cite{Strigari:2009bq,neutrinoBillard,neutrinoRuppin}. 

\par The low energy flux of atmospheric neutrinos, less than approximately 100 MeV, is difficult to directly measure and theoretically predict~\cite{Battistoni:2005pd}. At these energies, the uncertainty on the predicted atmospheric neutrino flux is approximately 20\%~\cite{Honda:2011nf}. Due to a cutoff in the rigidity of cosmic rays induced by the Earth's geomagnetic field at low energies, the atmospheric neutrino flux is larger for detectors that are nearer to the poles~\cite{Honda:2011nf}. 

\par Over all energies, the atmospheric neutrino flux peaks near the horizon, at zenith angle $\cos \theta \simeq 0$. At high energies, the flux is very nearly symmetric about $\cos \theta \simeq 0$, as at these energies the cosmic ray particles are more energetic than the rigidity cutoff. At low energies, the flux becomes asymmetric, as the flux of downward-going ($\cos \theta = 1$) neutrinos is lower than the flux of upward-going neutrinos ($\cos \theta = -1$). For the analysis in this paper, we consider the FLUKA results for the angular dependence of the atmospheric neutrino rate~\cite{Battistoni:2002ew}. As we discuss below, we find that when this flux is convolved with the angular dependence of the coherent neutrino-nucleus cross-section, the angular dependence is washed out and the recoil spectrum depends only weakly on direction. There is also a seasonal variation in the neutrino flux based on the atmospheric temperature which induces an additional time modulation. However the exact time dependence of this effect at the latitude of our mock experiment is not known and is likely too small to have a large effect on the observed limits. Hence for this study we ignore both the angular and time dependence of the atmospheric neutrino flux and model it as isotropic and constant in time,
\begin{equation}
  \frac{\textrm{d}^3 \Phi}{\textrm{d}E_\nu \textrm{d}\Omega_\nu \textrm{d}t} = \frac{1}{4\pi \Delta t}\frac{\textrm{d} \Phi}{\textrm{d} E_\nu}  \,.
\label{eq:ATMDSNBflux}
\end{equation}

\subsection{Diffuse supernova neutrinos}
\label{sec:dsnb}
\par For WIMP masses between 10 and 30 GeV, the neutrino floor is likely induced by the sub-dominant diffuse supernova neutrino background (DSNB), from all supernova explosions in the history of the Universe. The DSNB flux is a convolution of the core-collapse supernova rate as a function of redshift with the neutrino spectrum per supernova; for a recent review of the predicted DSNB flux see Beacom~\cite{Beacom:2010kk}. The DSNB spectra have a similar form to a Fermi-Dirac spectrum with temperatures in the range 3-8 MeV. We use the following temperatures for each neutrino flavour: $T_{\nu_e} =  3$ MeV, $T_{\bar{\nu}_e} = 5$ MeV and $T_{\nu_x} = 8$ MeV, where $\nu_x$ represents the four remaining neutrino flavours. Motivated by theoretical estimates we take a systematic uncertainty on the DSNB flux of 50\%. The DSNB is believed to be isotropic and constant over time, therefore its angular dependence can be expressed, as with the atmospheric neutrinos, using Eq.~(\ref{eq:ATMDSNBflux}).

\section{Neutrino and Dark Matter rate calculations}
\label{sec:rates}

\begin{table}[t]
\begin{ruledtabular}
\begin{tabular}{cccc}
$\mathbf{\nu}$ \bf{type} & $\mathbf{E_{\nu}^{\rm{max}}}$ \bf{(MeV)} & $\mathbf{E_{r_{\rm{Xe}}}^{\rm{max}}}$ \bf{(keV)} & $\mathbf{\nu}$ \bf{flux}\\
 & & & $\mathbf{(\rm{cm^{-2} \, s^{-1}})}$\\
\hline
$pp$ & 0.42341 & $2.94\times 10^{-3}$ & $\left(5.98\pm 0.006 \right) \times 10^{10}$\\
${}^{7}\rm{Be}$ & 0.861 & 0.0122 & $\left( 5.00\pm 0.07 \right) \times 10^9$\\
$pep$ & 1.440 & 0.0340 & $\left( 1.44\pm 0.012\right) \times 10^8$\\
${}^{15}\rm{O}$ & 1.732 & 0.04917 & $\left(2.23\pm 0.15\right) \times 10^8$\\
${}^{8}\rm{B}$ & 16.360 & 4.494 & $\left(5.58\pm 0.14\right) \times 10^6$\\
$hep$ & 18.784 & 5.7817 & $\left( 8.04\pm 1.30 \right) \times 10^3$\\
DSNB & 91.201 & 136.1 & $ 85.5\pm 42.7$\\
Atm. & 981.748 & $15.55\times 10^3$ & $10.5\pm 2.1$\\
\end{tabular}
\caption{Dominant neutrino fluxes with corresponding uncertainties. For the Solar neutrino flux, we utilize the high metallicity SSM, as described in the text. The maximum neutrino energy, $E_{\nu}^{\rm{max}}$, and maximum recoil energy on a Xenon target, $E_{r_{\rm{Xe}}}^{\rm{max}}$, are also shown.\label{tab:neutrino}}
\end{ruledtabular} 
\end{table}

\subsection{Coherent neutrino-nucleus elastic scattering}

\par We only consider the neutrino background from coherent neutrino-nucleus elastic scattering (CNS) as it produces nuclear recoils in the keV energy scale which cannot be distinguished from a WIMP interaction. We neglect neutrino-electron elastic scattering, mostly induced by $pp$ neutrinos, as it has been shown to only marginally affect the discovery potential of experiments with limited nuclear/electronic recoil discrimination for WIMP masses above 100 GeV~\cite{neutrinoBillard}.

\par Freedman~\cite{Freedman} has shown that neutrino-nucleon elastic scattering, which is well explained by the standard model but has yet to be observed, leads to a coherence effect at low momentum transfer that approximately scales with the atomic number of the target nucleus, $A$, squared. At higher recoil energies, generally above a few tens of keV, the loss of coherence is described by the nuclear form factor $F(E_r)$, for which we use the standard Helm form~\cite{lewin}. The differential cross-section as a function of the nuclear recoil energy ($E_r$) and neutrino energy ($E_\nu$) is given by 
\be
  \frac{\textrm{d} \sigma}{\textrm{d}E_r}(E_r,E_\nu) = \frac{G_F^2}{4 \pi} Q_W m_N \left(1-\frac{m_N E_r}{2 E_\nu^2} \right) F^2(E_r) \,,
\ee
where $Q_W = \mathcal{N} - (1-4\sin^2\theta_W) \mathcal{Z}$ is the weak nuclear hypercharge of a nucleus with $\mathcal{N}$ neutrons and $\mathcal{Z}$ protons, $G_F$ is the Fermi coupling constant, $\theta_W$ is the weak mixing angle and $m_N$ is the target nucleus mass. The directional and energy double differential cross-section can be written by noting that the scattering has azimuthal symmetry about the incoming neutrino direction so $\rm{d}\Omega_\nu = 2\pi \,\rm{d}\cos\beta$ and imposing the kinematical expression for the scattering angle, $\beta$, between the neutrino direction, $\hat{{\bf q}}_\nu$, and the recoil direction, $\hat{{\bf q}}_r $,
\be
 \cos\beta = \hat{{\bf q}}_r \cdot \hat{{\bf q}}_\nu = \frac{E_\nu + m_N}{E_\nu}\sqrt{\frac{E_r}{2 m_N}} \,,
\label{eq:kinematics}
\ee 
with $\beta$ in the range $(0,\pi/2)$, using a delta function,
\be
  \frac{\textrm{d}^2 \sigma}{\textrm{d}E_r \textrm{d}\Omega_r} = \frac{\textrm{d} \sigma}{\textrm{d}E_r} \, \frac{1}{2 \pi}\, \delta\left(\cos\beta - \frac{E_\nu + m_N}{E_\nu} \sqrt{\frac{E_r}{2 m_N}}\right) \,.
\label{eq:doublecrosssection}
\ee
The maximum recoil energy, $E_r^{\rm max}$, can be obtained by setting $\beta = 0$ in Eq.~(\ref{eq:kinematics}),
\be
E_r^{\rm max} = \frac{2 m_N E_\nu^2}{(E_\nu + m_N)^2} \approx \frac{2E_\nu^2}{m_N + 2E_\nu} \,.
\ee
The maximum recoil energies produced by the different types of neutrino for a Xenon target are shown in Table~\ref{tab:neutrino}.

\par The directional event rate per unit mass and time, as a function of the recoil energy, direction and time, is given by the convolution of the double differential CNS cross-section and the neutrino directional flux,
\be\label{eq:directionalrate}
  \frac{\textrm{d}^3 R}{\textrm{d}E_r \textrm{d}\Omega_r \textrm{d}t} =  \mathscr{N} \int_{E_\nu^{\rm min}} \frac{\textrm{d}^2 \sigma}{\textrm{d}E_r \textrm{d}\Omega_r}\times\frac{\textrm{d}^3 \Phi}{\textrm{d}E_\nu \textrm{d}\Omega_\nu \textrm{d}t} \textrm{d}E_\nu \textrm{d}\Omega_\nu \,,
\ee
where $E_\nu^{\rm min} = \sqrt{m_NE_r/2}$ is the minimum neutrino energy required to generate a nuclear recoil with energy $E_r$ and $\mathscr{N}$ is the number of target nuclei per unit mass.

 \begin{figure*}[t]
\begin{center}
\includegraphics[trim = 5mm 0 0mm 0mm, clip, width=1.0\textwidth,angle=0]{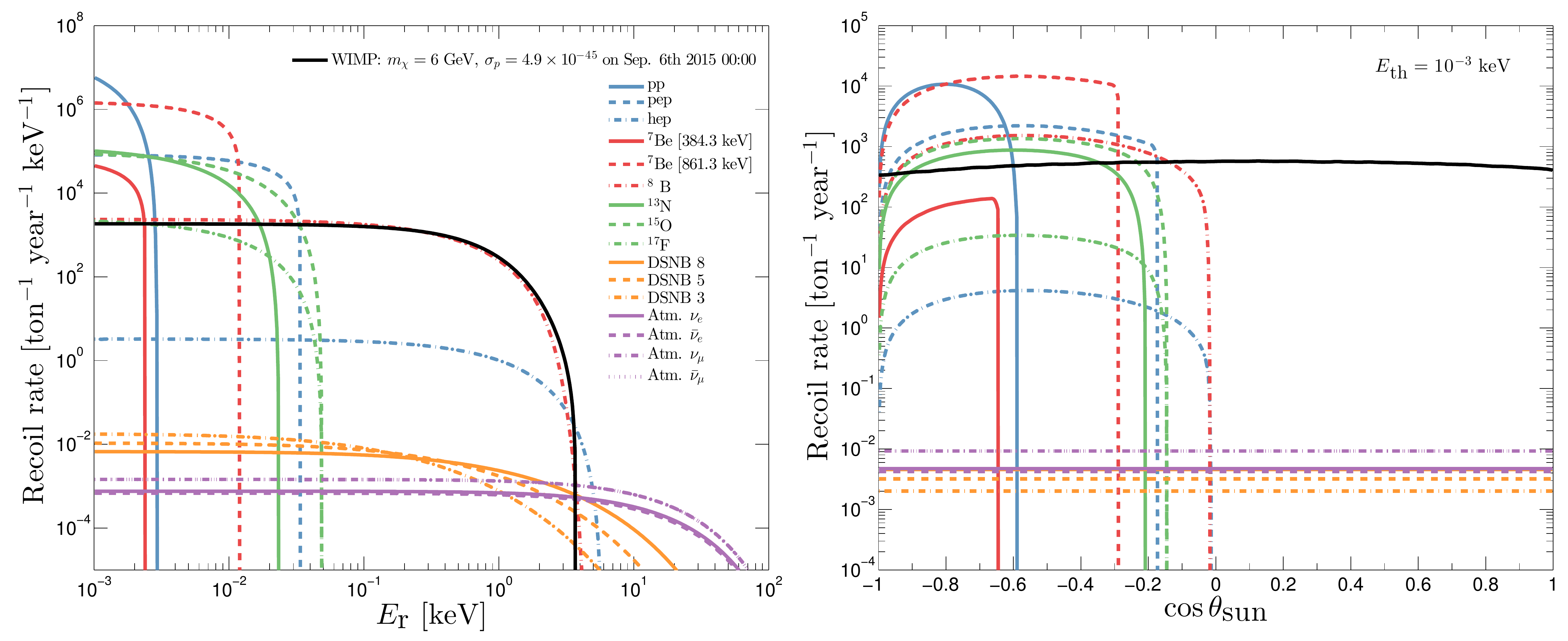}
\caption{Neutrino and $m_{\chi} = 6 \, {\rm GeV}$ WIMP nuclear recoil rates for a Xenon target as a function of recoil energy, $E_\textrm{r}$, (left) and cosine of the angle between the Solar vector and recoil vector, $\cos \theta_\textrm{sun}$, (right)  obtained by integrating the differential recoil spectrum over angle and energy respectively. The atmospheric and DSNB neutrino fluxes are taken to be isotropic. In order to show all types of neutrino the threshold energy has been set to 1 eV here.} 
\label{fig:neutrinorates}
\end{center}
\end{figure*} 

\par The directional event rate for Solar neutrinos is found by substituting Eqs.~(\ref{eq:solarneutrinoflux}), (\ref{eq:kinematics}) and (\ref{eq:doublecrosssection}) into Eq.~(\ref{eq:directionalrate}) and integrating over the neutrino direction $\Omega_\nu$,
\begin{align}
  &\frac{\textrm{d}^3 R}{\textrm{d}E_r \textrm{d}\Omega_r \textrm{d}t} = \frac{\mathscr{N}}{2\pi} \times \frac{1}{\Delta t}\left[ 1 + 2\epsilon\cos\left(\frac{2\pi(t- t_\nu)}{T_\nu}\right) \right] \times  \nonumber \\
& \int \frac{\textrm{d} \sigma}{\textrm{d}E_r}  \frac{\textrm{d} \Phi}{\textrm{d} E_\nu} \delta\left(\hat{{\bf q}}_r \cdot \hat{{\bf q}}_\odot - \frac{E_\nu + m_N}{E_\nu} \sqrt{\frac{E_r}{2 m_N}}\right) \textrm{d}E_\nu \,.
\label{eq:intermediate}
\end{align}
The delta function can then be rewritten as
\be
  \delta\left(\hat{{\bf q}}_r \cdot \hat{{\bf q}}_\odot - \frac{E_\nu + m_N}{E_\nu} \sqrt{\frac{E_r}{2 m_N}}\right) =  \frac{1}{E_\nu^\textrm{min}} \, \,\delta \left(x + \frac{1}{\mathcal{E}}\right) \, ,
\label{eq:delta}
\ee
where we have defined $x = -1/E_\nu$ and,
\be
  \frac{1}{\mathcal{E}} = \frac{\hat{{\bf q}}_r \cdot \hat{{\bf q}}_\odot}{E_\nu^\textrm{min}} - \frac{1}{m_N} \,.
\ee

Finally, by substituting Eq.~(\ref{eq:delta}) into Eq.~(\ref{eq:intermediate}), integrating over $x$ and converting back to $E_\nu$, we obtain an analytic expression for the directional event rate from Solar neutrinos,
\begin{align}
  \frac{\textrm{d}^3 R}{\textrm{d}E_r \textrm{d}\Omega_r \textrm{d}t} = & \frac{\mathscr{N}}{2\pi} \times \frac{1}{\Delta t}\left[ 1 + 2\epsilon\cos\left(\frac{2\pi(t- t_\nu)}{T_\nu}\right) \right]  \nonumber \\
 & \times \frac{\mathcal{E}^2}{ E_\nu^\textrm{min}} \frac{\textrm{d} \sigma}{\textrm{d}E_r}(E_r,\mathcal{E}) \, \, \frac{\textrm{d} \Phi}{\textrm{d} E_\nu}\bigg|_\mathcal{E} \,,
\label{eq:solarnu}
\end{align}
for $\cos^{-1}(\hat{{\bf q}}_r \cdot \hat{{\bf q}}_\odot) < \pi/2$ and 0 otherwise.

In the case of the atmospheric and diffuse supernova neutrinos, as we have assumed their fluxes to be isotropic and constant over time (see Sec.~\ref{sec:NeutrinoFluxes}), the directional event rate is simply given by substituting Eqs.~(\ref{eq:ATMDSNBflux}) and (\ref{eq:doublecrosssection}) into Eq.~(\ref{eq:directionalrate}) and integrating over the neutrino direction $\Omega_\nu$ leading to
\be
\frac{\textrm{d}^3 R}{\textrm{d}E_r \textrm{d}\Omega_r \textrm{d}t} = \frac{\mathscr{N}}{4\pi\Delta t} \int_{E_{\nu}^\textrm{min}} \frac{\textrm{d} \sigma}{\textrm{d}E_r}\times \frac{\textrm{d} \Phi}{\textrm{d}E_\nu } \textrm{d}E_\nu \,. 
\label{eq:isonu}
\ee

 \begin{figure}[t]
\begin{center}
\includegraphics[trim = 15mm 5mm 18mm 5mm, clip, width=0.49\textwidth,angle=0]{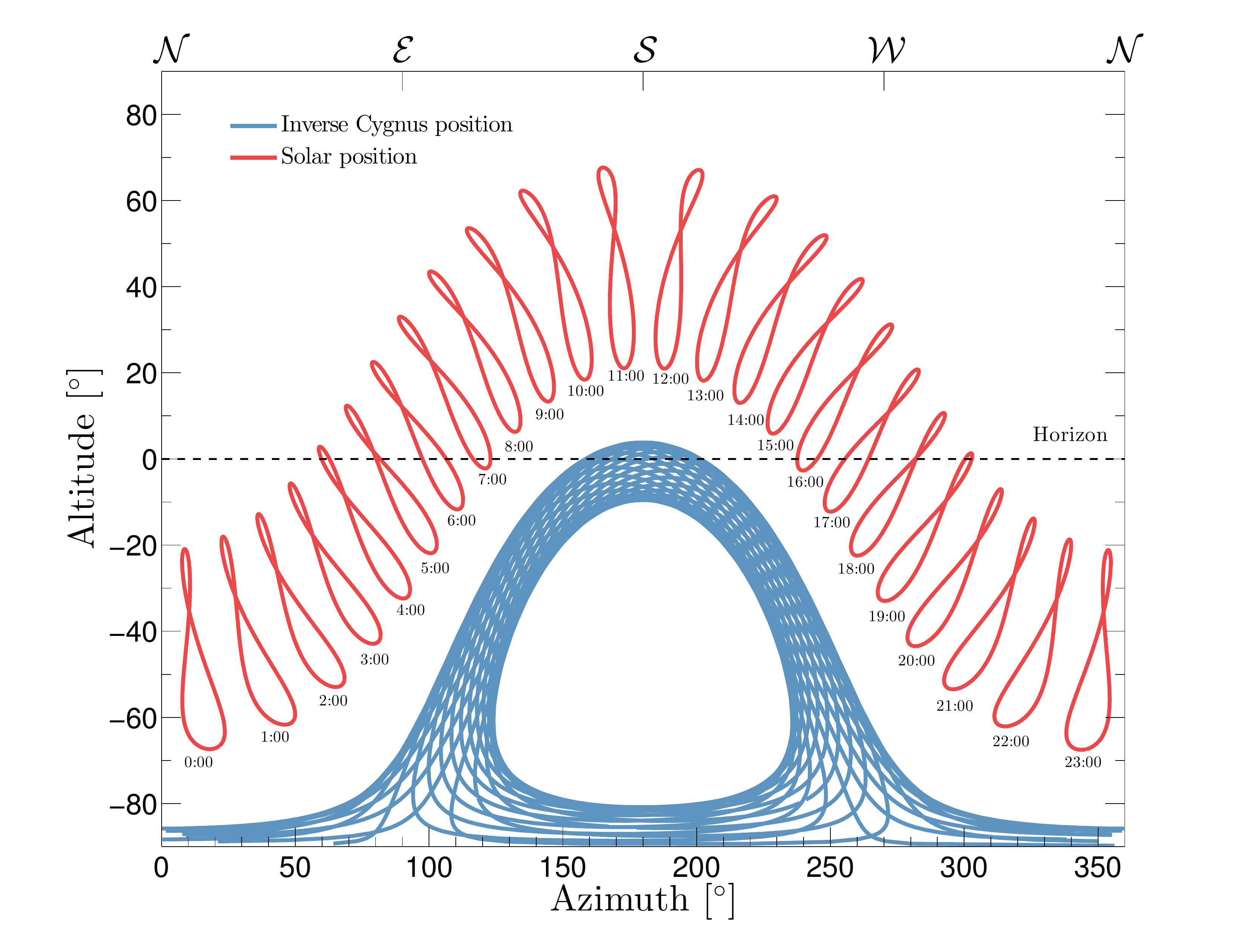}
\caption{The position in the sky, in terms of Altitude and Azimuth, of the Sun (red) and the inverse of the position of the constellation Cygnus (corresponding to the direction -$\bf{v}_\textrm{lab}$) (blue) as observed from the Modane underground laboratory (latitude $45.2^{\circ}$). The points show the position at measurements made every hour from the 1st January 2015 0:00 until 31st December 2015 23:00. The Solar position traces out 24 analemmas, corresponding to the Sun's position at each hour of the day over the course of a year. The dashed horizontal line is the horizon. As demonstrated here, the Sun's position does not coincide with that of Cygnus at any time.} 
\label{fig:solarcygnus}
\end{center}
\end{figure} 

The neutrino event rates as a function of energy and angle between Solar and recoil directions, $\cos{\theta_\textrm{sun}} = -\hat{\bf{q}}_r\cdot\hat{\bf{q}}_\odot$, obtained by integrating Eqs.~(\ref{eq:solarnu}) and ~(\ref{eq:isonu}) over direction and energy respectively are shown in Fig.~\ref{fig:neutrinorates}. Also shown is the recoil rate for a 6 GeV WIMP, showing the similarity between this spectrum and the spectrum of $^8$B neutrino recoils. The isotropic DSNB and atmospheric recoil rates are flat whereas the event rates of Solar neutrinos are highly anisotropic. The curves corresponding to the mono-energetic neutrinos ($^7$Be and $pep$) have a sharp cutoff in their directionality due to the finite energy threshold. From Fig.~\ref{fig:neutrinorates} one can already anticipate that the degeneracy between solar neutrino and WIMP events from an energy-only analysis will be almost completely removed with the addition of directional information.

\begin{figure*}[t]
\begin{center}
\includegraphics[trim = 0mm 0mm 0mm 0mm, clip, width=\textwidth,angle=0]{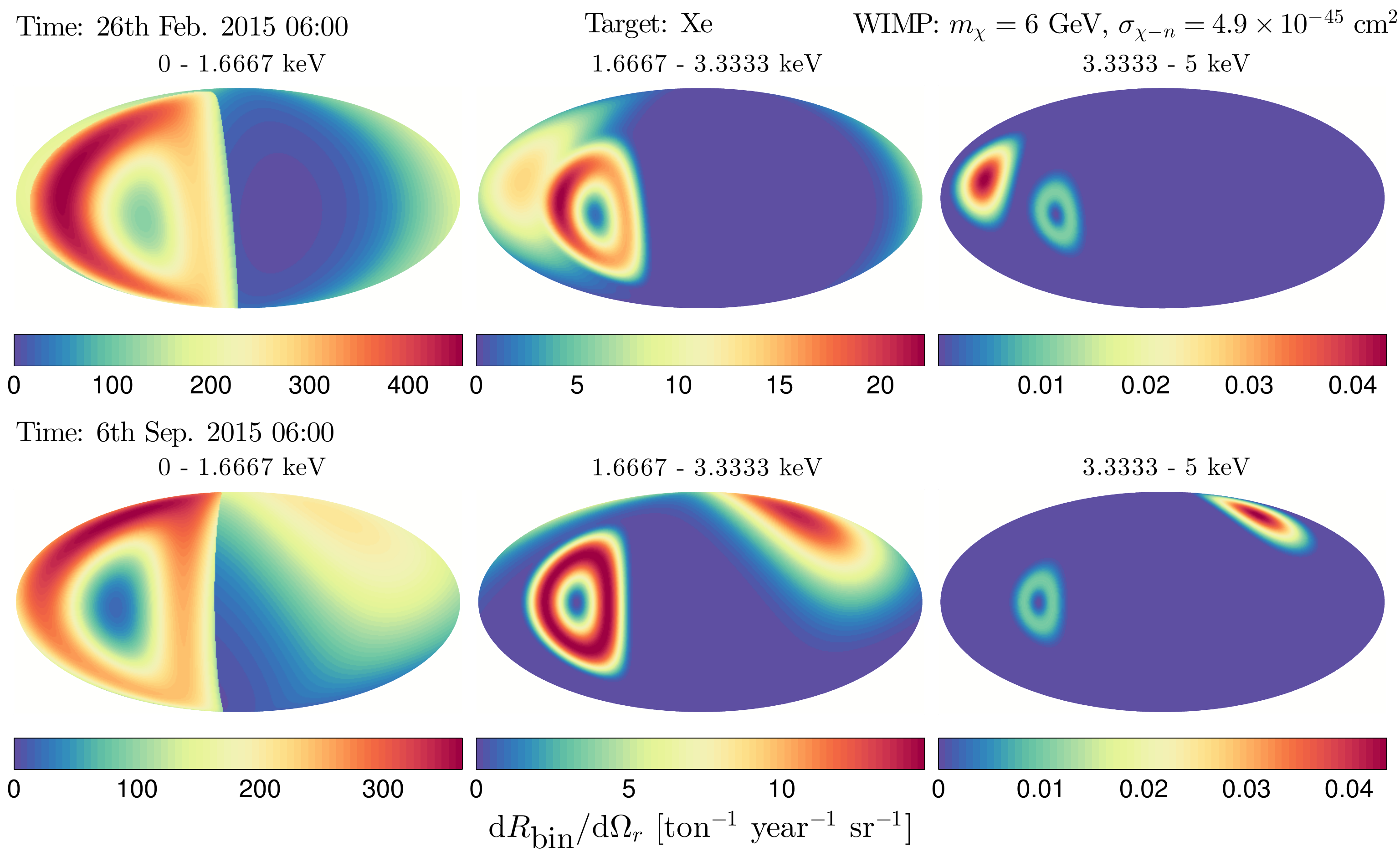}
\caption{Mollweide projections of the WIMP plus $^8$B neutrino angular differential event rate integrated within (from left to right) three equally sized energy bins spanning the range $E_{r} = 0 \,$ to $5 \, {\rm keV}$, for a WIMP with mass $m_{\chi} = 6 \, {\rm GeV}$ and $\sigma_{\chi-n} = 4.9 \times 10^{-45} \, {\rm cm}^2$ and a ${\rm Xe}$ target.
The top row shows the signal on February 26th, when the separation between the directions of the Sun and Cygnus is smallest ($\sim 60^{\circ}$), and the bottom row on September 6th, when the separation is largest ($\sim 120^{\circ}$). The WIMP contribution is to the left of the neutrino contribution on the top row and to the right on the bottom row. The Mollweide projections are of the event rate in the laboratory co-ordinate system with the horizon aligned horizontally and the zenith and nadir at the top and bottom of the projection respectively.} 
\label{fig:Moll}
\end{center}
\end{figure*} 

\subsection{Dark matter}

Like most spiral galaxies, the Milky Way is believed to be immersed in a halo of dark matter which outweighs the luminous component by at least an order of magnitude~\cite{salucci, klypin, Strigari:2013iaa}. The velocity distribution of dark matter in the halo is traditionally modelled (cf. Ref.~\cite{lewin}) as an isotropic Maxwell-Boltzmann distribution, corresponding to an isotropic halo with a $1/r^{2}$ density profile and a flat rotation curve. Simulated halos have velocity distributions which deviate systematically from the Maxwell-Boltzmann distribution~\cite{Vogelsberger:2008qb,Kuhlen:2009vh,Mao:2012hf,Nezri, Ling, Bruch, Read}. 
However, these deviations do not have a large effect on the discovery potential of directional detection experiments~\cite{Morgan:2004ys,Billard:2011zj}.
Therefore, and to allow comparison with previous work, we assume a truncated Maxwell-Boltzmann distribution throughout this work which in the Galactic rest frame has the form,
\begin{equation}
f_{\rm gal}({\bf v}) = \left\{
\begin{array}{rrll}
\rm & \frac{1}{N_{\rm esc}(2\pi\sigma^2_v)^{3/2}}\exp\left[-\frac{{\bf v} ^2}{2\sigma^2_v}\right] &	\ \text{if $|{\bf v}|<v_{\rm esc}$\,,}  \\
\rm & 0  		& 	\ \text{if $|{\bf v}|\geq v_{\rm esc}$\,,}
\end{array}\right.
\end{equation}
where $\sigma_v$ is the WIMP velocity dispersion which is related to the local circular speed, $v_0$, by $\sigma_v = v_0/\sqrt{2}$,  $v_{\rm esc}$ is the escape speed, and $N_{\rm esc}$ is a normalization constant,
\be
      N_{\textrm{esc}} = \textrm{erf}\left(\frac{v_{\textrm{esc}}}{\sqrt{2} \sigma_v}\right) - \sqrt{\frac{2}{\pi}} \frac{v_{\textrm{esc}}}{\sigma_v} \exp\left(-\frac{v_{\textrm{esc}}^2}{2 \sigma_v^2}\right) \, .
\ee
We use the conventional values for the circular and escape speeds: $v_0= 220  \, {\rm km \, s}^{-1}$~\cite{Kerr:1986hz} and $v_{\rm esc} = 544 \, {\rm km \, s}^{-1}$~\cite{Smith:2006ym}.

The velocity distribution of WIMPs in the rest frame of the laboratory is obtained through a Galilean transformation of the Galactic frame distribution, $f_{\rm gal}$, by the laboratory velocity ${\bf v}_{\rm lab}$ (discussed below),
\begin{equation}
f_{\rm lab}({\bf v}) =  f_{\rm gal}({\bf v} +{\bf v}_{\rm lab} ) \,.
\end{equation}

\begin{figure*}[t]
\begin{center}
\includegraphics[trim = 10mm 0 0mm 0mm, clip, width=0.99\textwidth]{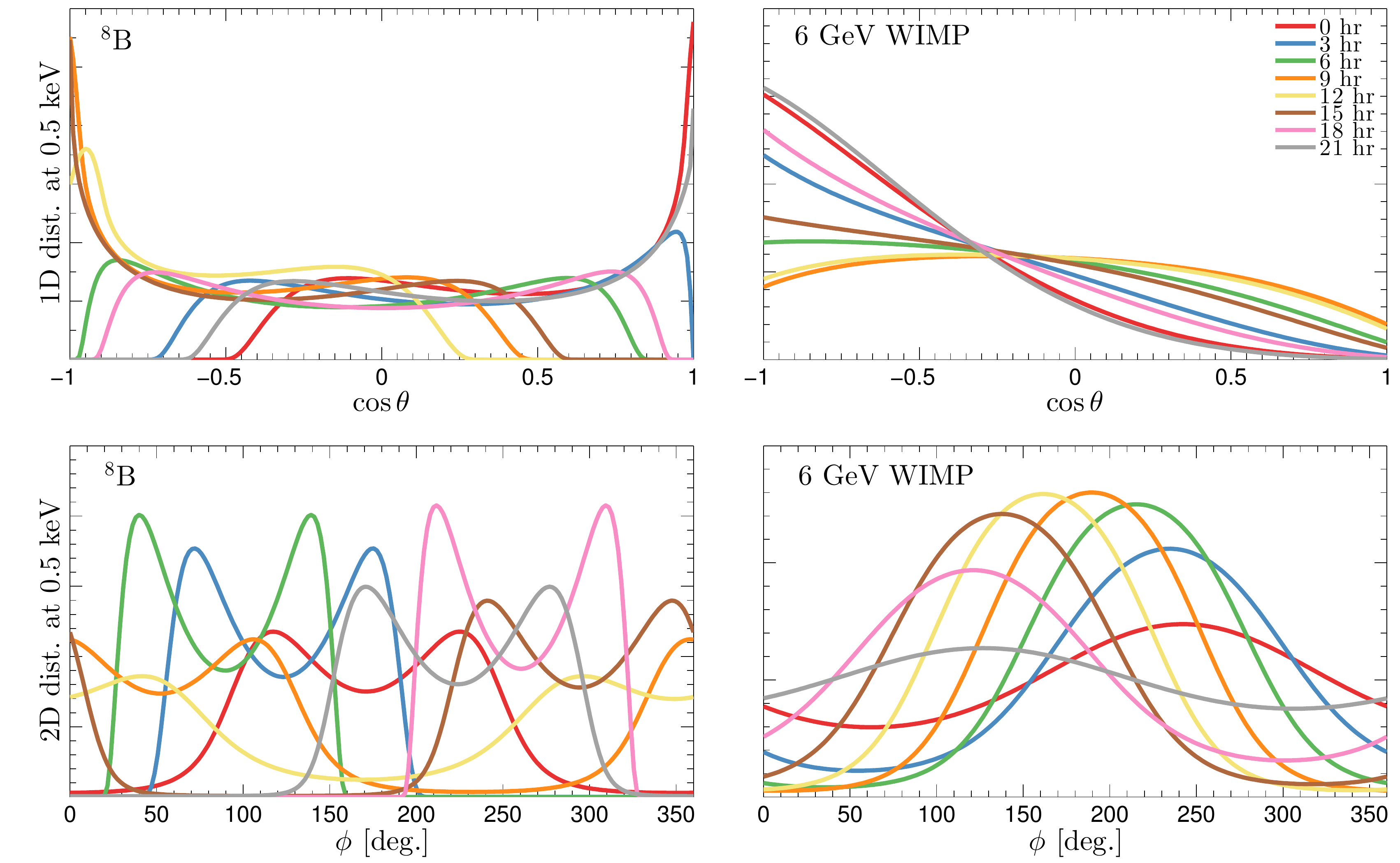}
\caption{The daily evolution, at three hourly intervals, of the angular distributions of 0.5 keV Xe recoils from
${}^{8}\rm{B}$ neutrinos (left column) and a WIMP with mass $m_{\chi} = 6 \, {\rm GeV}$ (right). The distributions are normalized to unity in each case and displayed with arbitrary units. The top row shows the distribution of $\cos{\theta}$  measured by a detector with 1-d readout and the bottom row the angle $\phi$  for  2-d readout. See the text for the definitions of $\theta$ and $\phi$. The date chosen was September 6th 2015, the date of maximum separation between the WIMP and neutrino distributions.} 
\label{fig:daily1d2d}
\end{center}
\end{figure*}

The directional event rate as a function of both recoil energy, direction in the lab frame and time, assuming spin-independent interactions with identical couplings to protons and neutrons, is given by~\cite{spergel,gondolo}
\begin{equation}
\frac{\mathrm{d}^3R}{\mathrm{d}E_r\mathrm{d}\Omega_r\mathrm{d}t} = \frac{\rho_0\sigma_{\chi-n}}{4\pi m_{\chi}\mu^2_{\chi n}\Delta t} A^2 F^2(E_r)\hat{f}_{\rm lab}(v_{\text{min}},\hat{\bf{q}}_{r}; t),
\label{eq:WIMPdirectionalrate}
\end{equation}
where $m_{\chi}$ is the WIMP mass, $\mu_{\chi n}$ the WIMP-nucleon reduced mass, $\rho_0 = 0.3$ GeV cm$^{-3}$  the local dark matter density, $A$ the mass number of the target, $\sigma_{\chi-n}$ the WIMP-nucleon cross-section and $v_{\text{min}} = \sqrt{2 m_N E}/{2 \mu_{\chi n}}$ is the minimum WIMP speed required to produce a nuclear recoil of energy $E_r$.
Finally, $\hat{f}_{\rm lab} (v_{\text{min}},\hat{\bf{q}}_{r}; t)$ is the three-dimensional Radon transform of the lab-frame WIMP 
velocity distribution $f_{\rm lab}({\bf v})$, which for the Maxwell-Boltzmann distribution has the form~\cite{gondolo},
\begin{align}
\hat{f}_{\rm lab}(&v_{\text{min}},\hat{\bf{q}}_{r}; t) = \frac{1}{N_{\rm esc}(2\pi\sigma_v^{2})^{1/2}}\nonumber \\
& \times  \left[\exp{\left(-\frac{\left|v_{\text{min}} + \hat{\bf{q}}\cdot{\bf v}_{\rm lab}\right|^2}{2\sigma^2_v}\right)} - \exp{\left(-\frac{v_{\rm esc}^2}{2\sigma^2_v}\right)}\right] .
\end{align} 
As can be seen from this equation,  if $v_{\rm min}<v_{\rm lab}$ the directional rate is maximum for $\hat{\bf{q}}\cdot{\bf v}_{\rm lab} = -v_{\rm min}$ (i.e., in a ring~\cite{ring}), while otherwise the rate is maximum for $\hat{\bf{q}} = -{\bf v}_{\rm lab}$ (i.e., a dipole distribution~\cite{spergel}). The strong correlation between the recoil directions and the laboratory motion in the Galactic frame allows the unambiguous authentication of a WIMP signal~\cite{Annedisco, Billarddisco}. 

The lab velocity, ${\bf v}_{\rm lab}$, is given by the sum of the rotation of the Solar System around the Galactic center ${\bf v}_{\rm GalRot}$, the peculiar velocity of the Solar System with respect to the local standard of rest, ${\bf v}_{\rm Solar}$, the Earth's revolution around the sun, ${\bf v}_{\rm EarthRev}$, and the Earth's rotation, ${\bf v}_{\rm EarthRot}$. The dominant contribution to the lab velocity is the sum ${\bf v}_{\rm GalRot} + {\bf v}_{\rm Solar}$ while ${\bf v}_{\rm EarthRev}$ and ${\bf v}_{\rm EarthRot}$ are, respectively, responsible for the annual~\cite{annual} and diurnal~\cite{daily} modulation effects. Even though their contributions to ${\bf v}_{\rm lab}$ are small, we take them into account in order to accurately study the additional discrimination power that can be brought by the annual and diurnal modulation effects. A detailed review of the Galactic-to-lab frame velocity and coordinate transformations are given in Ref.~\cite{daily}. In the following, we will consider a detector with the $\hat{{\bf x}}$,  $\hat{{\bf y}}$, and $\hat{{\bf z}}$ axes pointing toward the North, the West and the Zenith directions respectively. For a detector located at a latitude $90^\circ$, i.e., at the North Pole, $\hat{{\bf z}}$ is aligned with the spin axis of the Earth. Finally, in the detector frame, the direction of a recoil is given by the angles $\theta$ and $\phi$ defined such that,
\begin{equation}
\hat{{\bf q}} = \sin\theta\cos\phi \, \hat{{\bf x}} + \sin\theta\sin\phi \, \hat{{\bf y}} + \cos\theta \, \hat{{\bf z}} \,.
\label{eq:angles}
\end{equation}

The laboratory velocity $\bf{v}_\textrm{lab}$ happens to point towards the constellation of Cygnus. Figure~\ref{fig:solarcygnus} shows the position in the sky of the Sun and the {\it inverse} of the position of Cygnus (i.e., $-\bf{v}_\textrm{lab}$), as observed from the Modane underground laboratory (latitude $45.2^{\circ}$). The points show the position at observations made every hour from the 1st January 2015 0:00 until 31st December 2015 23:00. The Solar position traces out 24 analemmas, corresponding to the Sun's position at each hour of the day over the course of a year. As we see here, the Sun's position does not coincide with that of Cygnus at any time suggesting that a directional experiment should in principle be able to disentangle the WIMP from the Solar neutrino contributions in the observed data. As a matter of fact, the angular separation between the peak WIMP direction and the peak neutrino direction undergoes a sinusoidal modulation over the course of the year that varies from $60^{\circ}$ in February to $120^{\circ}$ in September.

\subsection{Resulting signals}
\label{sec:signals}
Figure~\ref{fig:Moll} shows Mollweide projections of the laboratory frame angular differential event rate from a 6 GeV WIMP plus $^8$B Solar neutrinos,  at the times when the separation between the directions of the Sun and Cygnus are smallest ($60^{\circ}$) and largest ($120^{\circ}$). This Figure clearly shows that, even at the time of smallest separation, the WIMP and neutrino recoil distributions can be easily distinguished as long as the angular resolution is better than a few tens of degrees. Although this Figure only shows the rates for $^8$B neutrino induced recoils, the angular distributions for other Solar neutrinos are very similar as neutrinos can only induce a recoil with an angle in the range $(0,\pi/2)$ from their incident direction. Additionally the angular dependence of the recoil spectra is correlated with energy as can be seen going from the left to the right hand panels. For both the WIMP and neutrino recoils the angular spread decreases with increasing energy i.e., the highest energy recoils have the smallest angle between the incoming particle direction and the recoil direction.

In addition to the standard case of a detector with full 3-dimensional sensitivity, we will also assess the discovery potential of a detector which only has sensitivity to 1-dimensional and 2-dimensional projections of the 3-d recoil track. Using Eq.~(\ref{eq:angles}) we define the 2-d readout to be the projection of the recoil track onto the $x$-$y$ plane such that only the angle $\phi$ is measured, and the 1-d readout to be the projection on to the $z$-axis such that only the angle $\theta$ is measured. 

Figure~\ref{fig:daily1d2d} shows the daily evolution of the 1-d, $\cos{\theta}$, and 2-d, $\phi$, recoil angle distributions at a single energy (0.5 keV) from ${}^{8}\rm{B}$ neutrinos and a WIMP with mass $m_{\chi} = 6 \, {\rm GeV}$. The $\phi$ distributions from $^8$B neutrinos have two peaks, because at a fixed recoil energy the neutrino energy spectrum produces recoils in a ring around the incident direction. In the WIMP case, however, the distribution of recoils is peaked in a single direction, towards $-\textbf{v}_\textrm{lab}$. The 2-d and 1-d distributions for both atmospheric and DSNB neutrinos are flat, and therefore we do not show them for clarity. The WIMP and neutrino distributions are significantly different, not only in their shape at a single time but also how they evolve over the course of a day. This suggests that a detector with only 1-d or 2-d readout should still be able to discriminate WIMP and neutrino induced recoils.

\section{Analysis Methodology} 
\label{sec:analysis}
In this section we introduce the analysis methodology we use to assess the discovery potential of each readout strategy for future low-threshold, ton-scale experiments. Discovery limits were first introduced in Ref.~\cite{Billard:2011zj} and are defined such that if the true WIMP model lies above this limit then a given experiment has a 90\% probability to achieve at least a 3$\sigma$ WIMP detection. To derive these limits, it is necessary to compute the detection significance associated with different WIMP parameters, for each detector configuration. This can be done  using the standard profile likelihood ratio test statistic~\cite{cowan2} where the likelihood function at a fixed WIMP mass is defined as
\begin{align}
\mathscr{L}(\sigma_{\chi-n},\boldsymbol{\Phi}) &= \frac{e^{-(\mu_\chi+\sum_{j=1}^{n_{\nu}}\mu^j_\nu)}}  {N!}\nonumber \\
& \times \prod_{i=1}^{ N}\left[\mu_\chi f_\chi({\bf q}_i,t_i) + \sum_{j=1}^{n_{\nu}}\mu^j_\nu f^j_\nu({\bf q}_i,t_i)\right]\nonumber \\
&\times \prod_{k=1}^{n_{\nu}}\mathscr{L}_k(\Phi_k) \,,
\end{align}
where the sums $j$ and $k$ are over the $n_\nu$ neutrino backgrounds,  $\mu_\chi$, $\mu^j_\nu$ and $N$ are, respectively, the expected number of WIMP and neutrino events, and the total number of observed events, $f_\chi$ and $f_\nu^j$ are the normalized, time and momentum dependent,  event rates for the WIMP and neutrinos, $t_i$ is the time at which the event occurred and ${\bf q}_i$ corresponds to the set of observables for each event, which depends on the readout considered. For 3-d readout ${\bf q}_i= \{E_r, \theta, \phi \}$.  Finally, $\mathscr{L}_k(\Phi_k)$ are the individual likelihood functions associated with the flux $\Phi_k$ of each neutrino component. These individual likelihood functions are each parametrized as gaussian distributions with a standard deviation given by the relative uncertainty in the neutrino flux normalization as discussed in Sec.~\ref{sec:NeutrinoFluxes} (see Table~\ref{tab:neutrino}).

The profile likelihood ratio corresponds to a hypothesis test between the null hypothesis $H_0$ (background only) and the alternative hypothesis $H_1$ which includes both background and signal, incorporating systematic uncertainties in this case the normalization of the neutrino fluxes. As we are interested in the WIMP discovery potential of future experiments, we test the background only hypothesis, $H_0$, on simulated data and try to reject it using the following likelihood ratio,
\begin{equation}
\lambda(0) = \frac{\mathscr{L}(\sigma_{\chi-n} = 0,\hat{\hat{\boldsymbol{\Phi}}})}{\mathscr{L}(\hat{\sigma}_{\chi-n},\hat{\boldsymbol{\Phi}})},
\end{equation}
where  ${\hat{\boldsymbol{\Phi}}}$ and $\hat{\sigma}_{\chi-n}$ denotes the values of ${\boldsymbol{\Phi}}$ and $\sigma_{\chi-n}$ that maximize the unconditional $\mathscr{L}$ and  
$\hat{\hat{\boldsymbol{\Phi}}}$ denotes the values of ${\boldsymbol{\Phi}}$ that maximize $\mathscr{L}$ under the condition $\sigma_{\chi-n} = 0$, i.e., we are profiling over the parameters in  ${\boldsymbol{\Phi}}$ which are considered to be nuisance parameters. As discussed in Ref.~\cite{cowan2}, the test statistic $q_0$ is then defined as,
\begin{equation}
q_0 = \left\{
\begin{array}{rrll}
\rm & -2\ln\lambda(0)	&	\ \hat{\sigma}_{\chi-n} > 0 \,, \\
\rm & 0  		& 	\ \hat{\sigma}_{\chi-n} < 0. 
\end{array}\right.
\end{equation}
A large value for this statistic implies that the alternative hypothesis gives a better fit to the data,  i.e., that it contains a WIMP signal.
The $p$-value, $p_0$, of a particular experiment is the probability of finding a value of $q_0$ larger than or equal to the observed value, $q_0^{\rm obs}$, if the null (background only) hypothesis is correct:
\begin{equation}
p_0 = \int_{q_0^{\rm obs}}^{\infty} f(q_0|H_0) \, {\rm d} q_0, 
\end{equation}
where $f(q_0|H_0)$ is the probability distribution function of $q_0$ under the background only hypothesis. Following Wilk's theorem, $q_0$ asymptotically follows a $\chi^2$ distribution with one degree of freedom (see Ref.~\cite{cowan2} for a more detailed discussion) and therefore the significance $Z$ in units of standard deviation is simply given by $Z = \sqrt{q^{\rm obs}_0}$. The discovery limit for a particular input WIMP mass can then be found by finding the minimum cross-section for which 90\% of the simulated experiments have $Z \geq 3$.
In the following results we have used distributions from 5000 Monte-Carlo experiments.

This analysis methodology was first introduced by the XENON10 collaboration~\cite{Aprile:2011hx} and many experiments are now using similar likelihood approaches e.g.,  LUX~\cite{Akerib:2013tjd}, CDMS-II~\cite{CDMSJulien,Agnese:2014xye}, and CoGeNT~\cite{Aalseth:2014jpa}. This has become possible thanks to the construction of accurate background models derived from reliable simulations, as well as data-driven analysis techniques based on calibration data. The advantage of using likelihood analyses is that they can not only determine whether or not a dark matter interpretation to the data is preferred and a WIMP signal detected, but they can also measure or constrain the WIMP parameters themselves. Furthermore a likelihood analysis will maximize the sensitivity to the dark matter signal and obtain the best possible limits for a given experiment. 

\section{Results}\label{sec:res} 
\subsection{Detector configurations}

As the goal of this paper is to give a detailed overview of how the neutrino background will affect future direct detection experiments, we first give a brief description of the different readout strategies that we consider. Here, and throughout, we consider a Xe-based experiment located in Modane with latitude and longitude (45.2$^\circ$,~6.67$^\circ$), taking data over a duration $\Delta t = 1$~year. As mentioned above,  we assume that the reference frame of the detector is such that $\hat{\bf x}$, $\hat{\bf y}$, and $\hat{\bf z}$ are, respectively, pointing toward the North, West, and Zenith directions and that the $\theta$ and $\phi$ angles are defined as in Eq.~(\ref{eq:angles}). Following previous work dedicated to the comparison of different readout strategies in the context of an arbitrary background~\cite{Billard:2014ewa}, we consider 6 detector readout strategies:
\begin{itemize} \itemsep0em 
 \item 3-d directional readout, $\{E_r,\theta,\phi,t\}$
 \item 2-d directional readout, $\{E_r,\phi,t\}$
 \item 1-d directional readout, $\{E_r,\theta,t\}$
 \item No directional information $\{E_r,t\}$
 \item Event time only $\{t\}$
 \item Number of events only (i.e., a counting experiment)
\end{itemize}

The last two strategies correspond to detectors that can only measure the total number of events above some threshold. This is the case for bubble chamber experiments~\cite{Cushman:2013zza} that adjust their operating pressure to nucleate a single bubble from a nuclear recoil. The energy and time, $\{E_r,t\}$, strategy corresponds to the majority of current and ongoing direct detection experiments where the kinetic energy of the recoiling nuclei is obtained from measurements of the heat, ionization and/or scintillation energies deposited in the detector (see Ref.~\cite{Cushman:2013zza} for a recent review).

The 3-d directional readout, $\{E_r,\theta,\phi,t\}$, corresponds to the ultimate detector that measures and exploits all the information available in the WIMP recoils. Current directional experiments are using low-pressure gaseous Time Projection Chambers (TPCs) in order to obtain tracks from $\mathcal{O}(10)$ keV nuclear recoils that are a few mm long (see Ref.~\cite{Ahlen} and references therein). For 3-d sensitive directional detectors, the track is measured by sampling over time the 2-dimensional projection of the ionization-induced electron cloud on a pixelized anode. For a 2-d readout, $\{E_r,\phi,t\}$, which is usually based on CCD technology, the anode is not time-sampled and therefore only a 2-dimensional projection of the drifted electron cloud can be measured. A 1-d readout, $\{E_r,\theta,t\}$, only measures the projection of the recoil track along the drift direction. This could be done for example in a dual-phase Liquid Xe TPC experiment by looking at the ratio of the ionization and scintillation energies thanks to columnar recombination of the drifting electrons/ions~\cite{nygren}. However, this effect has yet to be confirmed and ongoing measurements by the SCENE collaboration have only found mild evidence for columnar recombination from keV-scale nuclear recoils~\cite{Cao:2014gns}.

\subsection{Comparing readout strategies}
\begin{figure*}[t]
\begin{center}
\includegraphics[trim = 5mm 0mm 22mm 0mm, clip,width=0.49 \textwidth,angle=0]{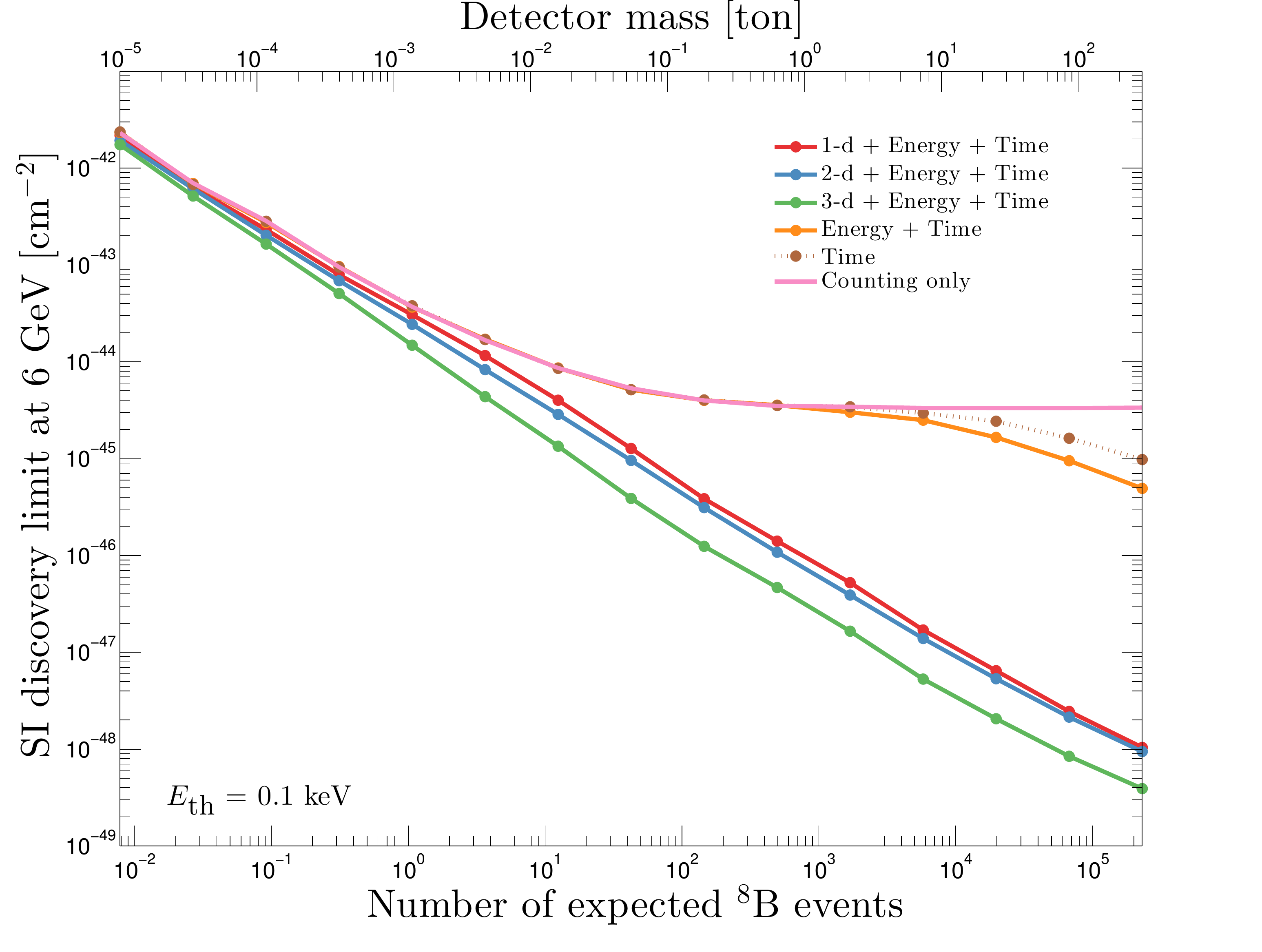}
\includegraphics[trim = 0mm 0mm 27mm 0mm, clip,width=0.49 \textwidth,angle=0]{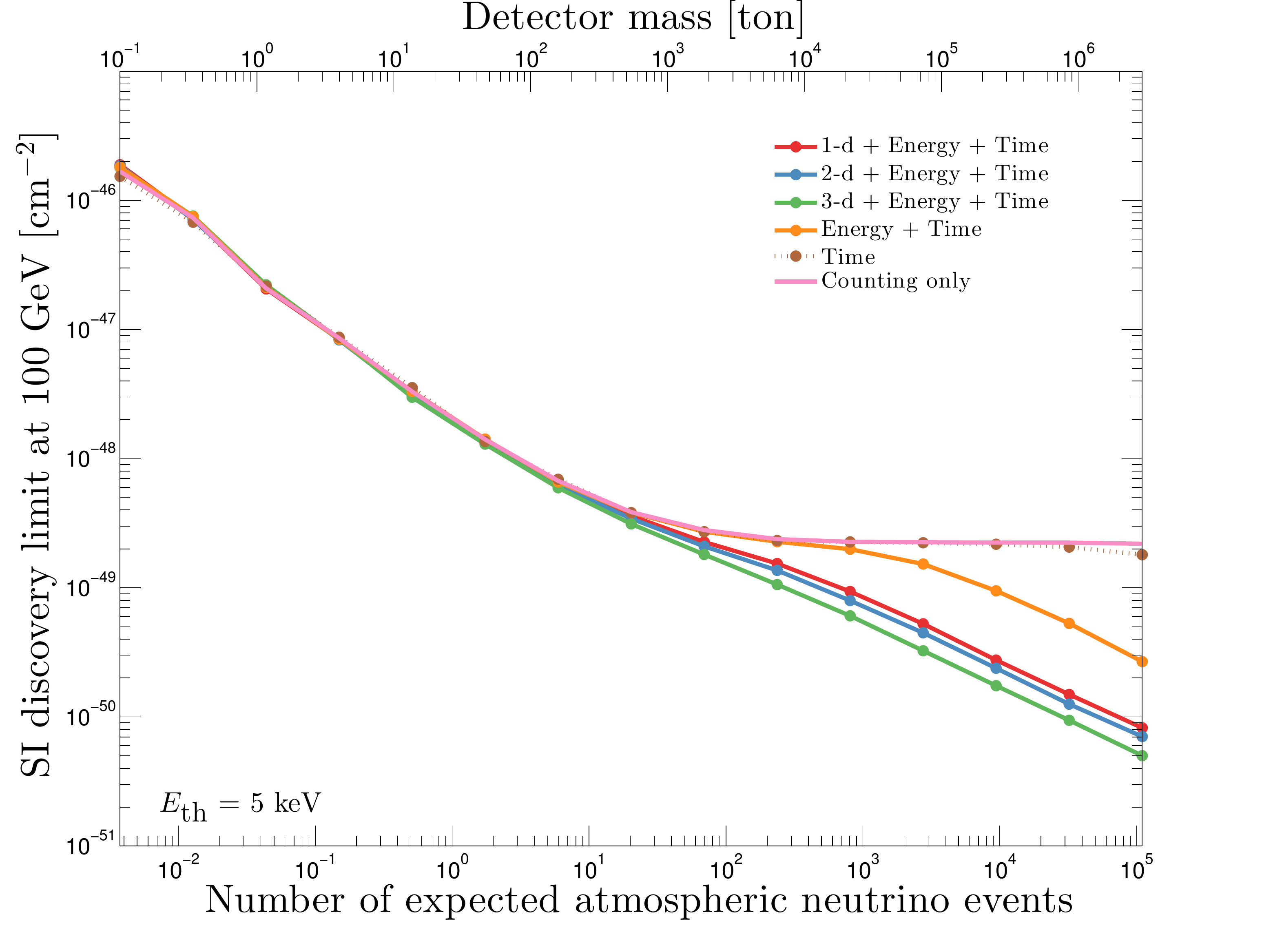}
\caption{The dependence of the discovery limit for the spin independent WIMP-nucleon cross-section, $\sigma_{\chi-n}$,  on the mass of a ${\rm Xe}$ detector operated for $1 \, {\rm year}$ using  (from top to bottom) 
the number of events only (pink line), time information (brown dotted),  energy \& time (orange), energy \& time  plus 1-d (red), 2-d (blue) and 3-d (green) directionality. 
The left (right) plot is for $m_{\chi} = 6 \, (100) \, {\rm GeV} $ and an energy threshold $E_{\rm th} = 0.1 \, (5) \, {\rm keV}$ and the bottom axis shows the number of ${}^{8}\rm{B}$  (atmospheric) neutrinos expected. 
Note the different scales of the left and right hand plots.} 
\label{fig:detectormass}
\end{center}
\end{figure*}

In this section we consider idealised detectors with perfect efficiency and angular resolution and no backgrounds apart for neutrinos, hence the discovery limits we obtain represent the best-case scenario for a particular readout strategy. We will study experimental limitations in Sec.~\ref{Sec:sense} and \ref{Sec:ang}. 

Figure~\ref{fig:detectormass} shows the evolution of the spin-independent discovery limit with detector mass, $M$, for a ${\rm Xe}$ detector taking data for one year using each of the different readout strategies. We consider two example WIMP masses and detector: a light WIMP \&  low threshold detector ($m_{\chi} = 6$ GeV and $E_{\rm th} = 0.1$ keV) and a 100 GeV WIMP \&  high threshold detector ($E_{\rm th} = 5 $ keV). For these two WIMP masses the recoil energy spectra closely matches those of $^8$B and atmospheric neutrinos respectively.  
As found previously, when the expected number of neutrino background events is negligible, the discovery limits improve rapidly with detector mass as a function of  $1/M$~\cite{neutrinoBillard} and the difference between the readout strategies is very small (cf. Ref.~\cite{Billard:2014ewa}). As the detector mass is increased and the experiment begins to have an appreciable neutrino background a Poisson background subtraction regime is entered and the discovery limit evolves as $1/\sqrt{M}$. When the expected number of neutrino events reaches $10-10^{2}$ the counting only, time only, and energy \& time limits plateau at a value controlled by the systematic uncertainty on the dominant neutrino component according to~\cite{neutrinoBillard},
\be
\sigma_{\rm DL} \propto \sqrt{\frac{1+\xi^2\mu_\nu}{\mu_\nu}},
\ee
where $\sigma_{\rm DL}$ is the discovery limit, $\xi$ is the uncertainty on the relevant neutrino flux (Table~\ref{tab:neutrino}) and $\mu_{\nu}$ is the expected number of neutrinos. In this regime the experiment cannot tell the difference between WIMP and neutrino induced recoils as there are not enough events to probe the different time dependences, or the differences in the tails of the energy distributions. This saturation of the WIMP sensitivity, which spans over two orders of magnitude in exposure, is what is commonly referred to as the neutrino floor.

The limits with directional readout however continue to decrease as the incorporation of directional information allows the distributions of WIMP and neutrino induced recoils to be distinguished. For the 100 GeV WIMP case, the limits from 2-d and 3-d readout are a factor of $\sim 1.2$ and $1.6$ better than those from 1-d readout whereas for the 6 GeV WIMP case they are factors of $\sim 1.2$ and $3$ times better. The discovery limit with directionality continues to decrease as $1/M$ for the 6 GeV WIMP as the directional and time-dependent distributions of the WIMP and Solar neutrino induced recoils have only very small overlap such that the background has very little effect on the discovery capabilities of the experiment. However for the 100 GeV WIMP the dominant background from isotropic atmospheric neutrinos significantly overlaps with the WIMP distribution so, although the experiment is able to distinguish the WIMP signal, the sensitivity is still compromised by the background. Therefore, in this case the discovery limit, beyond the saturation regime, evolves according to a standard Poisson background subtraction mode as $1/\sqrt{M}$. Our results are generally consistent with those of Grothaus {\it et al.}~\cite{Grothaus:2014hja}, in that we agree that directionality is the most promising strategy to go beyond the neutrino floor. However, there are quantitative differences. We find that directionality allows greater improvements in sensitivity than found in Ref.~\cite{Grothaus:2014hja}. This is largely due to the difference in energy thresholds used (2 keV in Ref.~\cite{Grothaus:2014hja} compared to 0.1 keV in this work) which drastically changes the ratio of WIMP to neutrino event numbers for low mass WIMPs. We also believe that some differences are a result of the fact that we have used the full directional information $\{\theta, \phi\}$, rather than just the reduced angles, and the two analyses also use different statistical techniques. We have hitherto considered an ideal detector, however finite angular resolution at the level considered in Ref.~\cite{Grothaus:2014hja} does not significantly change our conclusions (see Sec. \ref{Sec:ang}).

\begin{table*}[t]
\begin{ruledtabular}
\begin{tabular}{ccccccc}
{\bf Detector} & {\bf Target} & $\bf{M}$ {\bf (ton)} & $\mathbf{E}_{\rm{th}}$ {\bf (keV)} & $\Delta \mathbf{t}$ & {\bf (Lat., Long.)} & $\nu$ {\bf backgrounds} \\
\hline
A & Xe & 0.1 & 0.1 & 1 yr & ($45.2^{\circ}, 6.67^{\circ}$) & $^8$B, $hep$, Atm., DSNB \\
B & Xe & $10^4$ & 5 & 1 yr & ($45.2^{\circ}, 6.67^{\circ}$) & Atm., DSNB, $hep$ \\
\cite{neutrinoRuppin} (low) & Xe & 0.19 (ton-year) & 0.003 & - & - & $^7$Be, $^8$B, $pep$, $^{15}$O, $^{13}$N, $^{17}$F, $hep$, Atm., DSNB\\
\cite{neutrinoRuppin} (high)& Xe & $9.3\times 10^3$ (ton-year) & 4 & - & - & Atm., $hep$, DSNB\\
\end{tabular}
\caption{The properties of the two detector set-ups we consider: $M$ is the target mass, $E_{\rm th}$ the energy threshold and $\Delta t$ the length of time, from Jan. 1st 2015, over which data is taken. In both cases the detector is located at Modane. For reference we also show the properties of the detector set-ups used in Ref.~\cite{neutrinoRuppin} to generate the neutrino floor shown in Fig.~\ref{fig:readoutfloors} and subsequent plots. Also listed are the neutrino backgrounds present in each experiment, in order of number of expected events from highest to lowest.}
\label{tab:detectors}
\end{ruledtabular} 
\end{table*}

For very large detector masses ($M>10$ ton for $E_{\rm th} = 0.1$ keV and $M>10^{4}$ ton for $E_{\rm th} = 5 $ keV) which have accumulated more than $\sim 10^4$ neutrino events, the evolution of the Time only and Energy + Time discovery limits return to the Poisson background subtraction regime once more. With a very large number of events the time information allows discrimination between WIMP and neutrino induced recoils (cf. Ref.~\cite{Davis:2014ama}). However time information is more useful for discriminating Solar neutrinos from light WIMPs than for discriminating atmospheric neutrinos from heavier WIMPs. This is because the WIMP and Solar neutrino rates are both annually modulated, and also the amplitude of the annual modulation is larger for light WIMPs. For energy information only, with very large numbers of events the slight difference in the tails of the $^8$B neutrino and WIMP recoil energy distributions allows them to be discriminated~\cite{neutrinoRuppin}.

\begin{figure}
\begin{center}
\includegraphics[width=0.49\textwidth,angle=0]{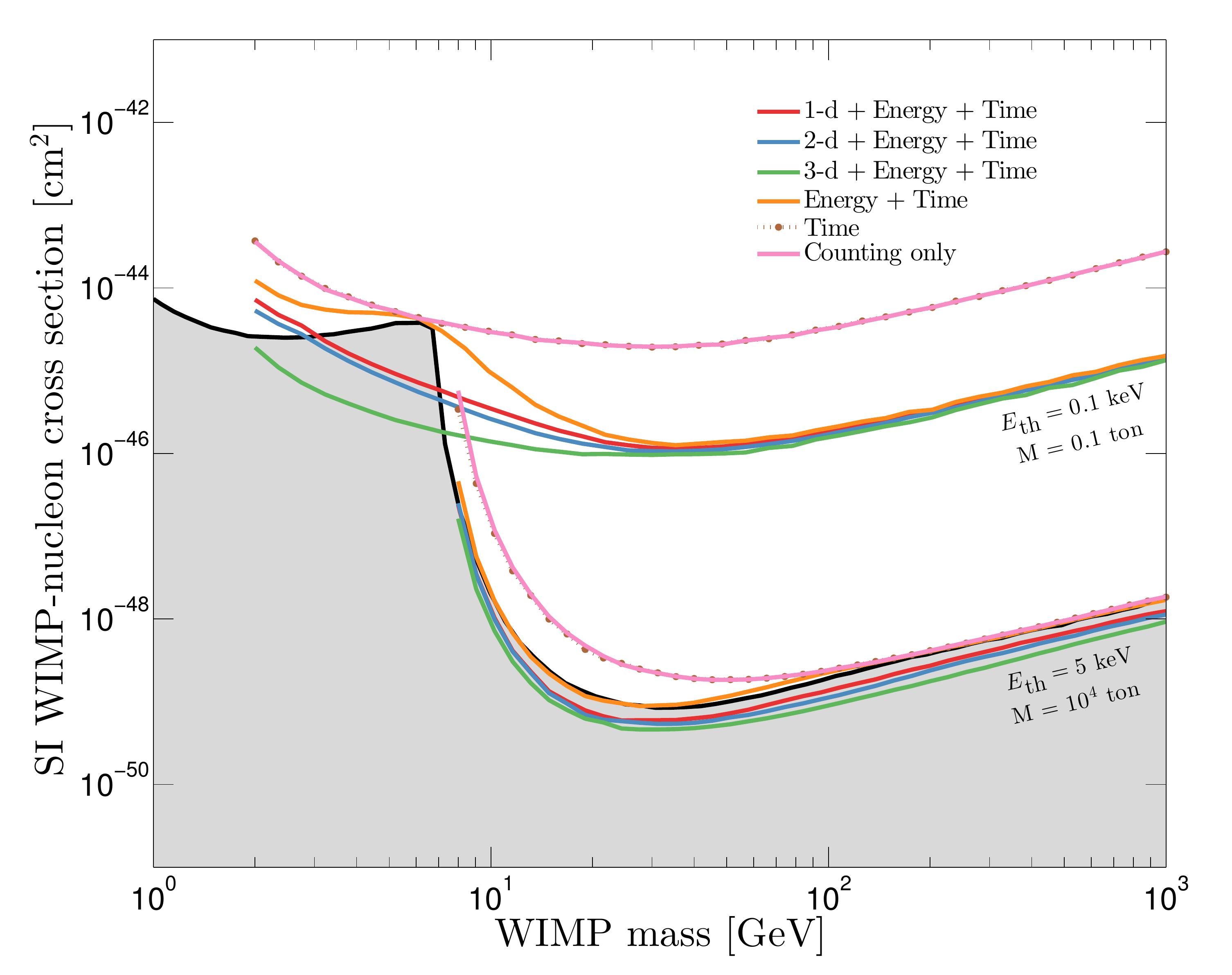}
\caption{%The discovery limit for the spin independent WIMP-nucleon cross-section, $\sigma_{\rm p}^{\rm SI}$, as a function of WIMP mass for a ${\rm Xe}$ detector operated for $1 \, {\rm year}$ 
The discovery limit
%, as in Fig.~\ref{fig:detectormass},
as a function of WIMP mass using (from top to bottom) 
the number of events only (pink solid line), time information (brown dotted),  energy \& time (orange), energy \& time  plus 1-d (red), 2-d (blue) and 3-d (green) directionality. 
The upper (lower) set of lines are for the two detector set-ups described in Table~\ref{tab:detectors}: Detector A (B) with a target mass $M=0.1 \, (10^4)$ ton and an energy threshold $E_{\rm th} = 0.1 \, (5) \, {\rm keV}$. 
The black curve and shaded region shows the neutrino floor from Ref.~\cite{neutrinoRuppin}.} 
\label{fig:readoutfloors}
\end{center}
\end{figure} 

Having studied the evolution of the discovery limit as a function of detector mass for two specific WIMP masses, we now consider two fixed example detector set-ups outlined in Table~\ref{tab:detectors}: a low mass \& low threshold detector ($M=0.1$ ton and $E_{\rm th} = 0.1 \, {\rm keV}$ respectively) and a high mass \& high threshold detector ($10^4$ ton and 5 keV). Again, for simplicity and to probe the full annual modulation signal, we assume that data is accumulated over one year. These detector masses and thresholds are chosen so that a non-directional detector with the same mass and threshold would be in the saturation regime that results in the neutrino floor, as seen in Fig.~\ref{fig:detectormass}.

Figure~\ref{fig:readoutfloors} shows the discovery limit as a function of WIMP mass for the two detector set-ups for each readout strategy. Also shown as the shaded region is the neutrino floor from Ref.~\cite{neutrinoRuppin} which is the combination of two limits obtained by a Xenon detector. For light WIMPs ($m_{\chi} <10$ GeV) the limit comes from a 3 eV threshold detector with an exposure of 0.19 ton years, while for heavier WIMPs ($m_{\chi} >10$ GeV) a detector with a 4 keV threshold and an exposure of $9.3 \times 10^3$ ton years was used.
The two detector configurations roughly match our two detector setups A and B in Table~\ref{tab:detectors}. As described in Refs.~\cite{neutrinoBillard, neutrinoRuppin}, the low-mass part of the neutrino floor comes from solar neutrinos (which have low-energies but high fluxes) with the shoulder at $m_\chi = 6$~GeV arising due to $^8$B neutrinos. The high-mass part, above $\sim 10$ GeV, is due to DSNB neutrinos and atmospheric neutrinos which have higher energies but much lower fluxes.

For the low threshold detector, the directional discovery limits clearly cut through the low-mass neutrino floor and for the 3-d readout there is actually almost no reduction in sensitivity due to the neutrino background. The 1-d and 2-d readouts do suffer a small reduction in sensitivity, but evidently the distributions are different enough that it is still possible to probe cross-sections below the limit set by non-directional experiments. For the high threshold detector the improvement in the discovery limits, with respect to the high-mass neutrino floor, from directionality is smaller. However it does still help discriminate the isotropic atmospheric neutrino background from WIMP induced recoils, in particular for WIMP masses around $100 \, {\rm GeV}$ where the energy spectra from WIMPs and atmospheric neutrinos are most similar.

In summary, we found that directionality is a powerful tool for disentangling neutrino backgrounds from a putative WIMP signal. The gain from directionality is particularly impressive for low mass WIMPs thanks to the large separation between the solar neutrino and WIMP incoming directions, see Sec.~\ref{sec:signals}. Interestingly, we found that this result still holds even if only the 2-d or 1-d projection of the recoil tracks can be measured. The gain from directionality in the high-mass region is more moderate, however, due to the large overlap between the WIMP and the isotropic DSNB and atmospheric neutrino distributions. Even in this case, we found that 1-d and 2-d readouts still outperform non-directional experiments. This highlights that it is worthwhile to construct directional detectors, even without full 3-d readout.
Similar conclusions were reached in Ref.~\cite{Billard:2014ewa} in the context of a discriminating WIMPs from an arbitrary isotropic background.

\subsection{Sense recognition}
\label{Sec:sense}
\begin{figure}[t]
\begin{center}
\includegraphics[width=0.49\textwidth,angle=0]{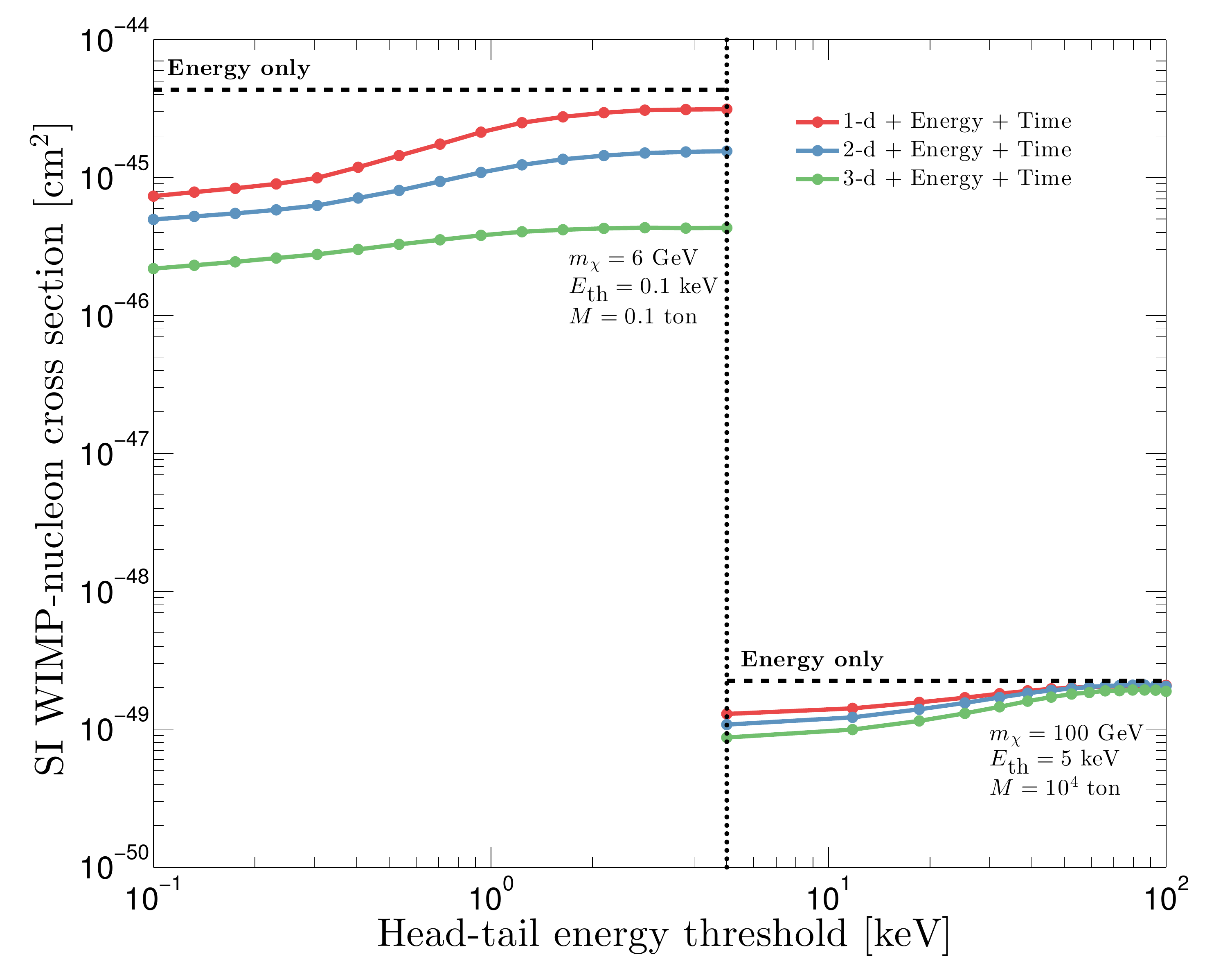}
\caption{
The discovery limit
%, as in Fig.~\ref{fig:detectormass}, 
as a function of the energy threshold for head-tail discrimination for 1-d (red), 2-d (blue) and 3-d (green) directional readout (with energy and time information in all three cases). The dashed black line shows the discovery limit with energy information only.
The left (right) hand set of curves is for $m_{\chi} = 6 \, (100) \, {\rm GeV} $ and detector set-up A (B), with low (high) mass and threshold.}
%a detector with a target mass $M=0.1 \, (10^4)$ ton and an energy threshold $E_{\rm th} = 0.1 \, (5) \, {\rm keV}$.} 
\label{fig:headtail}
\end{center}
\end{figure} 

\begin{figure}[t]
\begin{center}
\includegraphics[width=0.49\textwidth,angle=0]{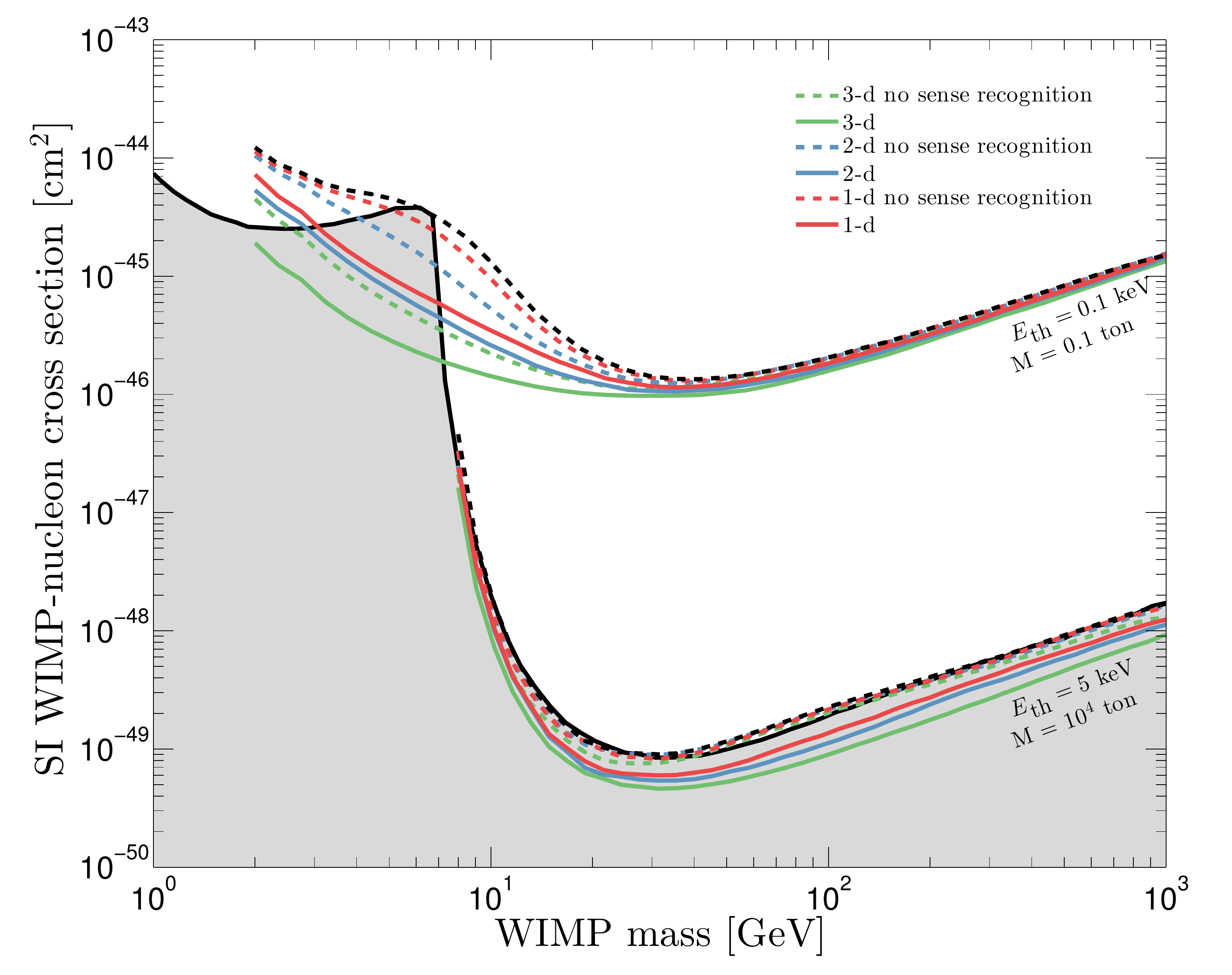}
\caption{%The discovery limit for the spin independent WIMP-nucleon cross-section, $\sigma_{\rm p}^{\rm SI}$, as a function of WIMP mass for a ${\rm Xe}$ detector. 
The discovery limit
%, as in Fig.~\ref{fig:detectormass}, 
as a function of WIMP mass for detectors with full sense recognition (solid lines) and no sense recognition (dashed) and 1-d (red), 2-d (blue) and 3-d (green) directional readout. The upper (lower) set of lines are for detector set-up A (B), with low (high) mass and threshold.  The dashed black lines show our discovery limit with energy information only and
the black curve and shaded region shows the neutrino floor from Ref.~\cite{neutrinoRuppin}.}
\label{fig:headtailfloors}
\end{center}
\end{figure} 

\begin{figure*}[t]
\begin{center}
\includegraphics[trim = 5mm 0mm 22mm 0mm,clip,width=0.49 \textwidth,angle=0]{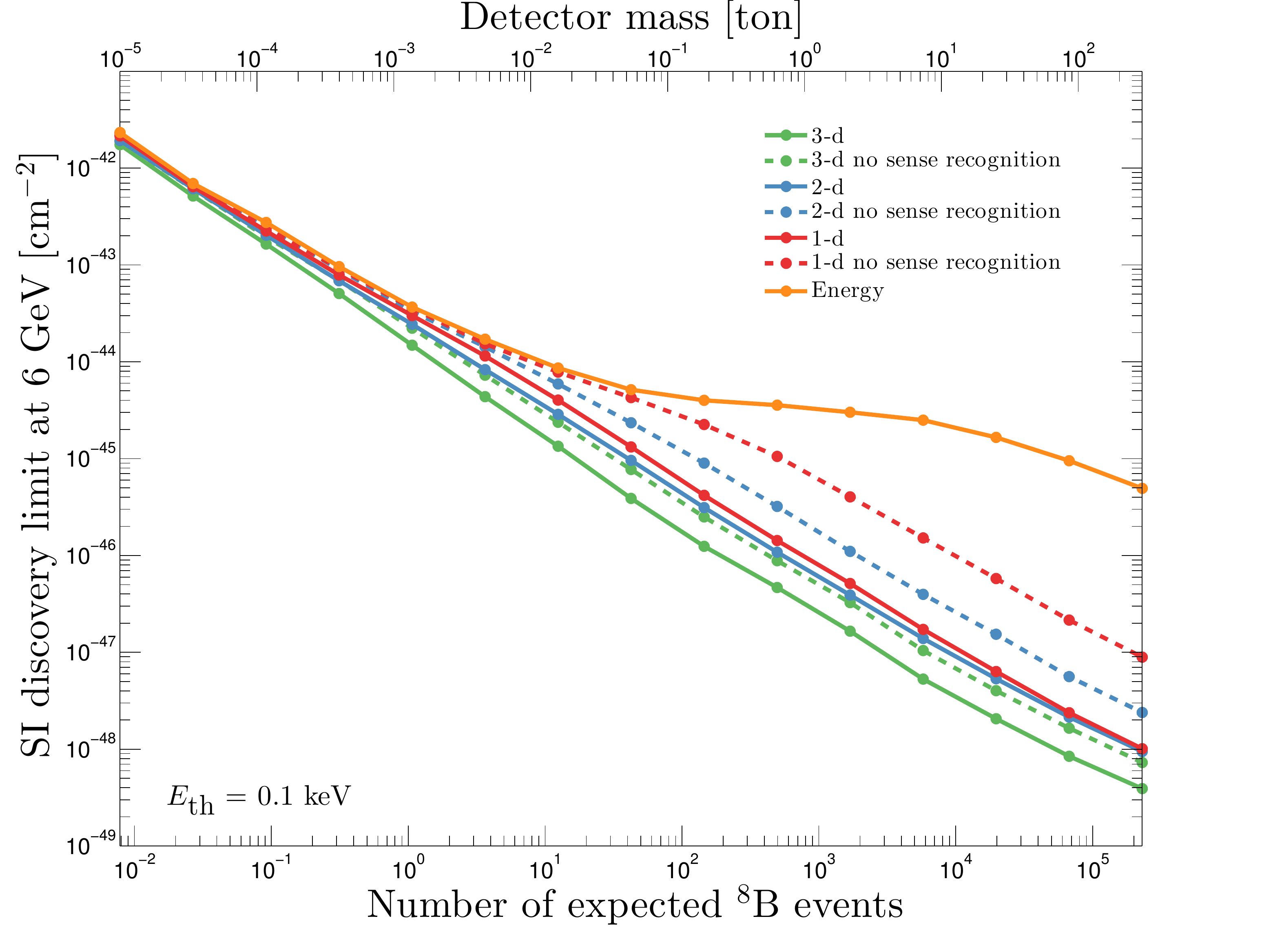}
\includegraphics[trim = 0mm 0mm 27mm 0mm,clip,width=0.49 \textwidth,angle=0]{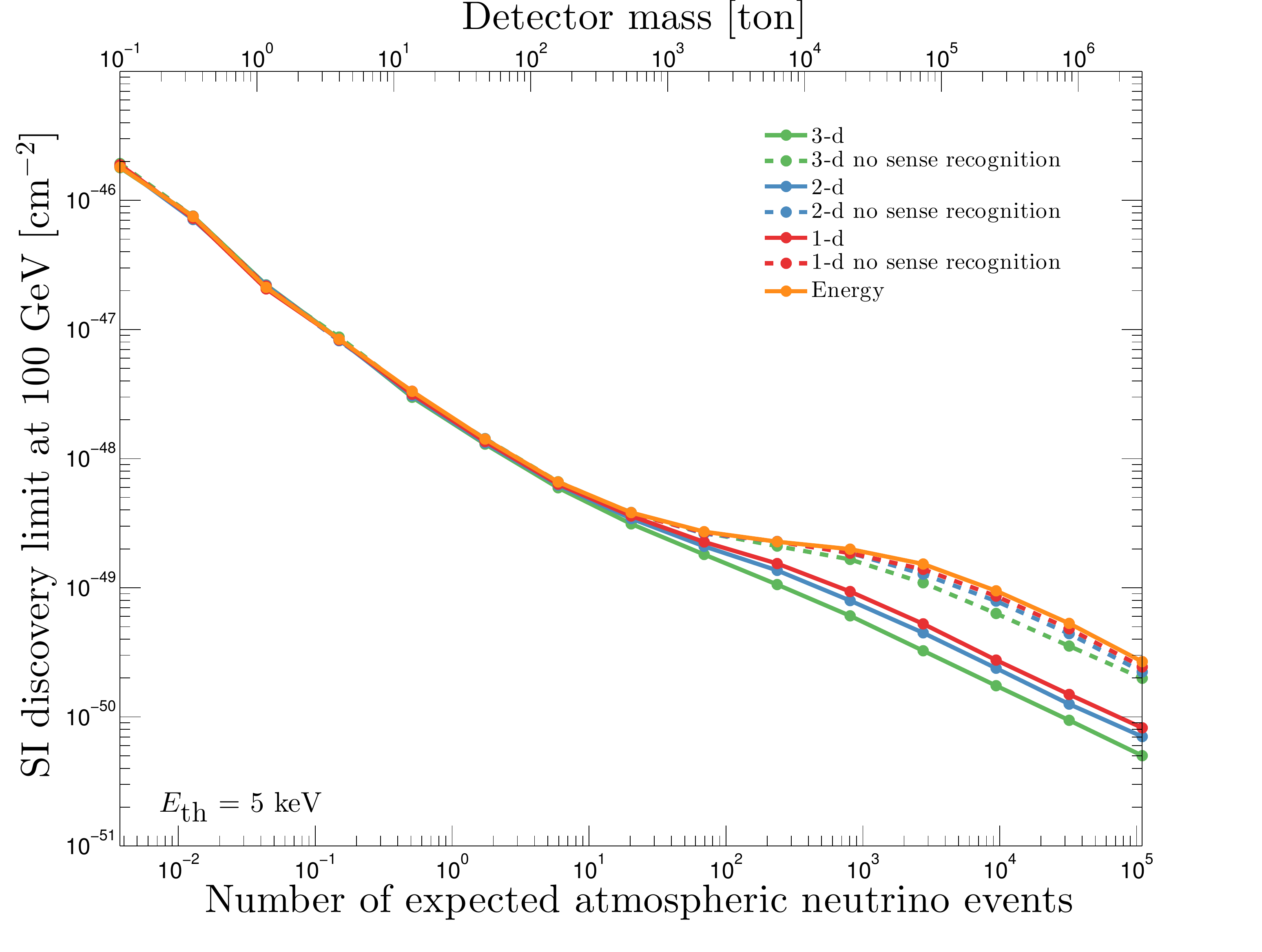}
\caption{The dependence of the discovery limit for the spin independent WIMP-nucleon cross-section, $\sigma_{\chi-n}$, on detector mass for detectors with (solid lines) and without (dashed lines) sense recognition for each of the three directional readout strategies 3-d (green), 2-d (blue) and 1-d (red). The discovery limit for energy + time only readout is shown with the orange lines. The left (right) panel is for 
$m_{\chi} = 6 \, (100) \, {\rm GeV} $ and detector set-up A (B), with low (high) mass and threshold.\label{fig:headtailmass}}
%The dependence is shown for the same two WIMP mass cases as in Fig.~\ref{fig:detectormass}, left: 6 GeV (with a 0.1 keV threshold), and right: 100 GeV (with a 5 keV threshold).} 
\end{center}
\end{figure*}

As well as the 3-dimensional reconstruction of the recoil track, a key experimental concern is head-tail recognition i.e., the ability to measure the sense ($+ \hat{{\bf q}} $ or $- \hat{{\bf q}}$) of the nuclear recoils. Determining the sense of a recoil is expected to be possible in a gas TPC by measuring the asymmetry in the charge collected along the recoil track as well as asymmetry in the shape of the track itself~\cite{Billard:2011kv}. Whilst there has been important progress with the experimental effort, sense recognition remains one of the outstanding challenges for current and future detectors~\cite{Majewski:2009an, Dujmic:2008zz}. Whether or not this information can be retrieved from a detector plays a key role in its ability to discriminate WIMP induced recoils from backgrounds~\cite{Morgan:2004ys,Billard:2011zj,Copi:2005ya}. This is because in the absence of head-tail discrimination the angular recoil rates in the forward and background directions are added together and the anisotropy of the WIMP induced recoils is effectively decreased.

Figure~\ref{fig:headtail} shows how the discovery limit depends on the energy threshold for head-tail discrimination, for 
$m_{\chi} = 6$ and $100 \, {\rm GeV} $ and our two example detector set-ups (Table~\ref{tab:detectors}) with 3-d, 2-d and 1-d readouts. 
For simplicity, we assume that above (below) the head-tail energy threshold there is perfect (no) head-tail discrimination.
For the light WIMP and the low mass and threshold detector, the discovery limits are weakened as the head-tail energy threshold is increased from $0.1 \, {\rm keV}$ to $\sim 1-2 \, {\rm keV}$ before flattening off to a factor between $\sim 1.5$ (1-d)  and $\sim 10$ (3-d) below the energy only limit. For lower dimensional readout the decrease in sensitivity is larger and the plateau in the limit is reached for a larger head-tail energy threshold. Qualitatively similar behaviour occurs for the $100 \, {\rm GeV} $ WIMP and the high mass and threshold detector. In this case the discovery limits flatten off to values $1.1 - 1.2$ below the
energy only limit at a head-tail energy threshold of $60 \, {\rm keV}$. 

In Figure~\ref{fig:headtailfloors} we show the discovery limits with and without sense recognition, as a function of WIMP mass. The factor by which the discovery limit changes without sense recognition is largest for light, $m_{\chi} < {\cal O}(20 \, {\rm GeV})$, WIMPs and a low threshold. The discovery limit achieved by a 3-d readout is still considerably lower than the non-directional limit however 1-d and 2-d readouts do suffer without sense recognition and are only marginally better than the non-directional limits, especially at high WIMP masses.

In Figure~\ref{fig:headtailmass} we show (in similar fashion to Fig.~\ref{fig:detectormass}) the evolution of the discovery limit now as a function of detector mass for $m_{\chi} = 6$ and $100 \, {\rm GeV} $ with and without sense discrimination.
As in Fig.~\ref{fig:headtailfloors}, we see that the lack of sense recognition is most damaging in the 100 GeV WIMP case. This is particularly true for the 1-d and 2-d readouts where with no sense recognition there is only a factor of 1.1 and 1.2 improvement over a detector with no directional information at all and the evolution of the discovery limit suffers from the same saturation effect due to the similarity in the recoil distributions. In the 6 GeV WIMP case the discovery limits with no sense recognition continue to decrease past the saturation regime suffered by the non-directional limit. However there is still a reduction in sensitivity by factors of 1.9, 2.8 and 8.9 for 3-d, 2-d and 1-d readouts respectively compared to the limits with sense recognition. Interestingly, the discovery limit for 3-d readout with no sense recognition is slightly better than 1-d and 2-d readouts with sense recognition.

Our main conclusion regarding sense recognition is that for discriminating between Solar neutrinos and low mass WIMPs, having 3-d readout with no method of determining sense is marginally preferable to 1-d or 2-d readout with sense determination. 
This is because the recoil distributions from low mass WIMPs and Solar neutrinos are both anisotropic and have sufficient 3-d angular separation that they are still distinguishable even without recoil sense information. For the higher mass WIMPs this is not the case, and 
without sense recognition the advantage of directionality is almost entirely lost, even in 3-d.

\subsection{Angular resolution}
\label{Sec:ang}

The final experimental limitation we study is that of finite angular resolution caused by the inaccuracy in the estimation of the ``true'' recoil direction. This is an inherent difficulty faced by all directional detectors.  For instance, directional detectors using low pressure gas TPCs suffer from straggling effects as the recoiling nucleus collides with other gas nuclei and more importantly from the diffusion of the primary electrons while drifting toward the anode~\cite{Billard:2011kv}. Finite angular resolution will smear out the WIMP and Solar neutrino distributions in Fig.~\ref{fig:Moll}, making it more difficult to discriminate between the two. Since the minimum separation between the peak WIMP and neutrino directions is $\sim60^\circ$, an angular resolution better than this will likely be required to differentiate between the WIMP and Solar neutrino distributions.

Finite angular resolution results in a recoil in the direction $\hat{r}'(\Omega'_r)$ being reconstructed in the direction $\hat{r}(\Omega_r)$ with a probability distribution that has the form of a Gaussian smoothing kernel on a sphere~\cite{Copi:2005ya,Billard:2011zj}
\be
K(\Omega_r,\Omega'_r) = \frac{1}{ (2\pi)^{3/2} \sigma_\gamma \rm{erf}(\sqrt{2}\sigma_\gamma)   } \exp{\left( - \frac{\gamma^2}{2\sigma_\gamma^2} \right)} \,,
\ee
where $\gamma$ is the angle between the original and reconstructed directions,
\be
\cos{\gamma} = \sin{\theta}\sin{\theta'}\cos{(\phi-\phi')} + \cos{\theta}\cos{\theta'} \,,
\ee
in the co-ordinates defined in Eq.~\ref{eq:angles}. The measured directional recoil rate is then the convolution of the smoothing kernel with the original directional recoil rate
\be
\frac{{\rm d}^2 R}{\rm{d} \Omega_r \rm{d}E_r} = \int_{\Omega'_r} \frac{{\rm d}^2 R}{\rm{d} \Omega'_r \rm{d}E_r}(\Omega'_r, E_r) K(\Omega_r,\Omega'_r) \, {\rm d} \Omega'_r \,.
\ee
\begin{figure}[t]
\begin{center}
\includegraphics[width=0.49\textwidth,angle=0]{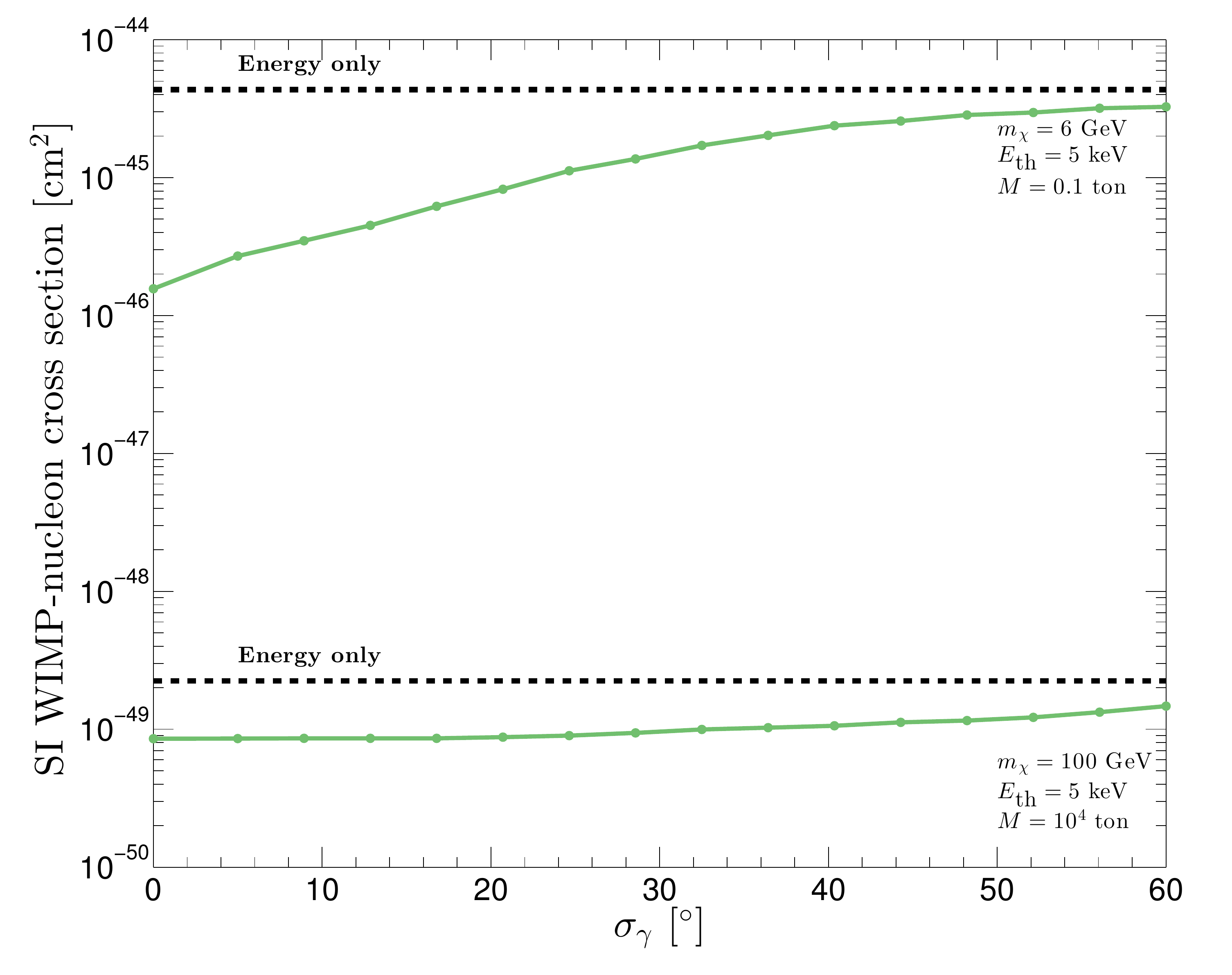}
\caption{%The discovery limit for the spin independent WIMP-nucleon cross-section, $\sigma_{\rm p}^{\rm SI}$ as a function of angular resolution $\sigma_\gamma$.  
The discovery limit
%, as in Fig.~\ref{fig:detectormass}, 
as a function of angular resolution, $\sigma_\gamma$ for a detector with 3-d readout.  
The upper (lower) set of lines are for $m_{\chi} = 6 \, (100) \, {\rm GeV}$ and detector set-up A (B), with low (high) mass and threshold.  
The dashed lines show the discovery limit using energy information only. }
\label{fig:angressingle}
\end{center}
\end{figure} 

\begin{figure}[t]
\begin{center}
\includegraphics[width=0.49 \textwidth,angle=0]{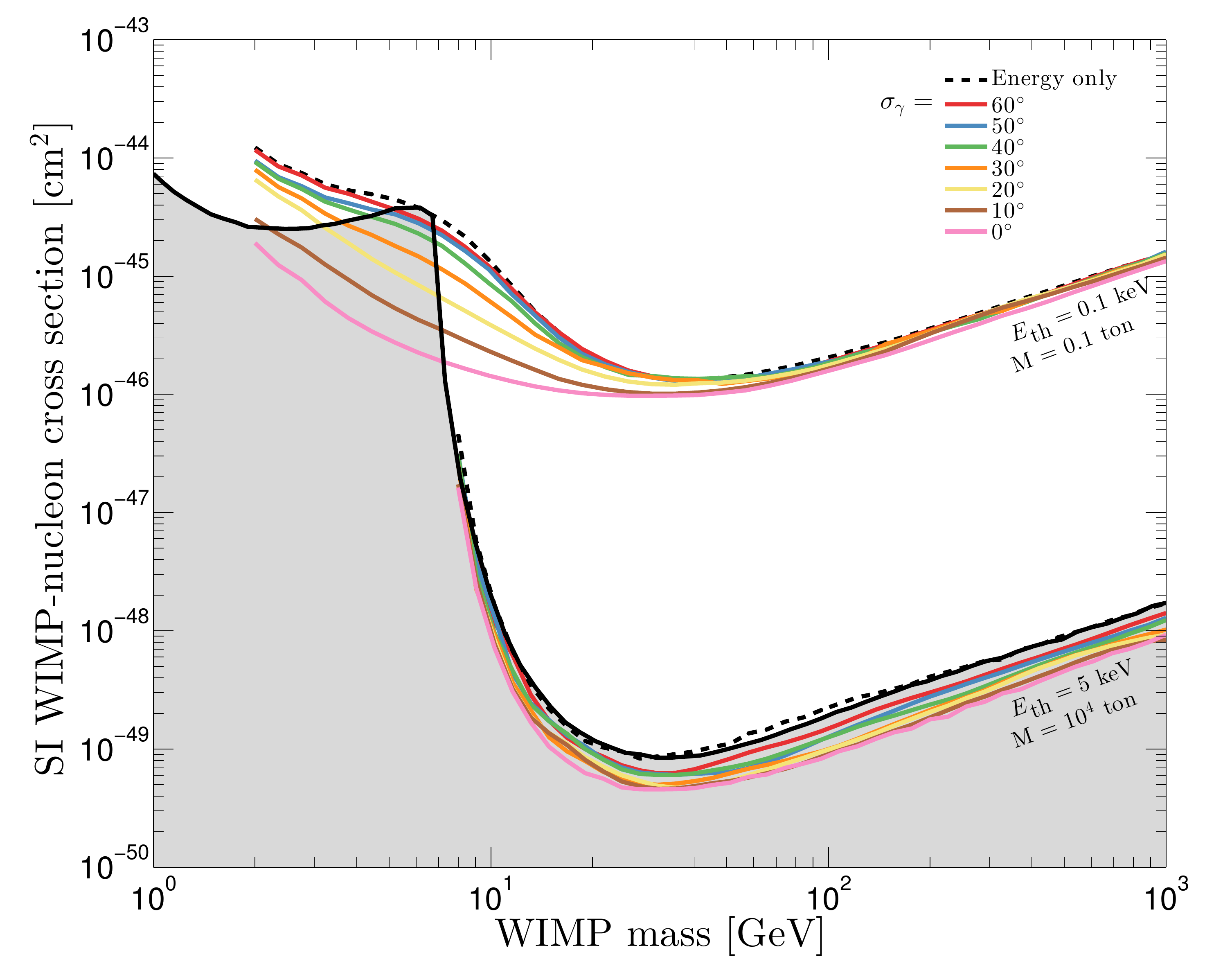}
\caption{%The discovery limit for the spin independent WIMP-nucleon cross-section, $\sigma_{\rm p}^{\rm SI}$, for a ${\rm Xe}$ detector as a function of $m_\chi$ for a ${\rm Xe}$ detector with angular resolution, $\sigma_\gamma$, varying between $0^{\circ}$ and $60^{\circ}$. 
The discovery limit, as in Fig.~\ref{fig:readoutfloors}, as a function of WIMP mass with angular resolution, $\sigma_\gamma$, varying between $0^{\circ}$ and $60^{\circ}$ for a detector with 3-d readout. 
The upper (lower) set of lines are for detector set-up A (B), with low (high) mass and threshold.  
The dashed black lines show our discovery limit with energy information only and
the black curve and shaded region shows the neutrino floor from Ref.~\cite{neutrinoRuppin}. 
} 
\label{fig:angresfloors}
\end{center}
\end{figure} 

The discovery limit as a function of angular resolution, $\sigma_\gamma$, is shown in Fig.~\ref{fig:angressingle} for
$m_{\chi} = 6$ and $100 \, {\rm GeV} $ for our two example detector set-ups (Table~\ref{tab:detectors}) with 3-d readout. As expected,  finite resolution makes it harder to discriminate a $6 \, {\rm GeV}$ WIMP from Solar neutrinos. The discovery limit is an order of magnitude weaker for $\sigma_{\gamma} = 30^{\circ}$ than for perfect angular resolution, and for $\sigma_{\gamma} > 50^{\circ}$ the limit is only marginally better than that obtained using energy information only. For the heavier WIMP and the more massive detector, the discovery limit only has a slight change with increasing $\sigma_\gamma$. This is because the finite angular resolution affects only the WIMP signal and not the isotropic background from atmospheric neutrinos. However in this case the improvement afforded with directionality, even in the ideal case, is smaller.

Figure~\ref{fig:angresfloors} shows the discovery limits for the two detector set-ups as a function of WIMP mass and angular resolution. Finite angular resolution significantly limits the ability of a low threshold directional detector to discriminate light, $m_{\chi} < {\cal O}(20 \, {\rm GeV})$, WIMPs from Solar neutrinos. The effects of finite angular resolution are greatest for $m_\chi \sim 6$ GeV, when the energy spectra of WIMPs and $^8$B neutrinos match one another. For the high threshold detector the reverse behaviour is observed. At higher WIMP masses the effect of increasing angular resolution is more apparent than at lower masses ($<12$ GeV), this is because the anisotropy of the recoil distribution decreases with increasing WIMP mass.

The main conclusion of this sub-section is that angular resolution of order $\sigma_{\gamma} = 30^{\circ}$ or better is required to exploit the different directional signals of light WIMPs and Solar neutrinos. For angular resolutions larger than this there is little benefit from having the directional information at all as the Solar neutrino and WIMP signals are poorly resolved. For heavier WIMPs the neutrino floor can still be overcome even with angular resolutions up to $60^\circ$. This is because the dipole asymmetry of the WIMP recoil distribution has a large dispersion and the effect of smearing due to finite angular resolution is less significant. Therefore for light WIMPs probing cross-sections below the $^8$B neutrino floor requires good angular resolution, however for the atmospheric neutrino floor the experimental limits can be competitive even with only modest angular resolution.

\section{Conclusions}
\label{sec:conc} 

We have studied in detail how direct detection experiments with directional sensitivity can subtract the background due to coherent neutrino-nucleus scattering and circumvent the so-called neutrino floor over a wide range of WIMP masses. In particular for light WIMPs (which have a similar recoil energy spectrum to $^8$B Solar neutrinos) directionality would allow a ton-scale low threshold detector to be sensitive to cross-sections several orders of magnitude below the neutrino floor. We have also shown that experiments that can only measure 1-d or 2-d projections of the recoil tracks can still discriminate WIMPs from neutrino backgrounds.

Moving beyond ideal detectors, we studied the effects of finite angular resolution and limited sense recognition.
The angular distributions of WIMP and Solar neutrino induced recoils are sufficiently different that for light WIMPs sense recognition is not crucial. The discovery limits are a factor of roughly two and ten worse without sense recognition for 3-d and 1-d readout respectively. However the discovery limits still improve strongly with increasing exposure. The discovery limit for 3-d readout with no sense recognition is slightly better than 1-d and 2-d readouts {\it with} sense recognition. For heavier WIMPs, however, sense recognition is required to discriminate WIMPs from the isotropic background from atmospheric neutrinos.
Finally we found that if the angular resolution is worse than of order thirty degrees, then it becomes significantly more difficult to discriminate between light WIMPs and Solar neutrinos.  Angular resolution is less crucial for distinguishing heavier WIMPs from isotropic atmospheric neutrino (although in this case the improvement offered by an ideal directional detector is smaller).

We have used Xenon as a target nucleus throughout, although we note that no directional dark matter detectors using this material currently exist (and the only proposed directional detection strategy using Xe has a 1-d readout~\cite{nygren}). Using Xenon allows our results to be easily compared with previous studies of the neutrino floor~\cite{neutrinoBillard,neutrinoRuppin,Davis:2014ama,Grothaus:2014hja} and simplifies the analysis as it is not necessary to consider the effects of multiple target nuclei, as is the case with CF$_4$ which is most commonly used in current low pressure gas TPCs~\cite{Ahlen}. Moreover, multiple target experiments and their complementarity have already been studied extensively in Ref.~\cite{neutrinoRuppin}. We have also not accounted for astrophysical uncertainties both in the values of parameters such as the escape speed and lab velocity but also in the shape of the velocity distribution. Ref.~\cite{Davis:2014ama} studied the effect on the light WIMP neutrino floor of a non-Maxwellian speed distribution and a distribution containing a stream. It found that the inclusion of time information allowed the additional uncertainty from the speed distribution to be overcome, and the neutrino floor suppressed. We expect that directionality would also help in a similar way.

The results presented in this paper make a compelling case for the development of large directional dark matter detectors.
If the results of the next generation of direct detection experiments lead the search to smaller WIMP-nucleon cross-sections, new techniques will need to be implemented to tackle the neutrino background. We have shown that the use of directionality is a powerful way of doing this, even for non-ideal detectors.\\

%{\bf Acknowledgments}
\begin{acknowledgments}
CAJO and AMG are both supported by the STFC, AMG also acknowledges support from the Leverhulme Trust. JB is grateful to the LABEX Lyon Institute of Origins (ANR-10-LABX-0066) of the Universit\'e de Lyon for its financial support within the program ``Investissements d'Avenir'' (ANR-11-IDEX-0007) of the French government operated by the National Research Agency (ANR).  LES  acknowledges  support from NSF grant PHY-1522717.
\end{acknowledgments}

\end{document}